\documentclass[aps,showpacs,preprint]{revtex4}
\usepackage{graphicx}
\usepackage{amssymb}

\begin{document}

\title{The elementary excitations of the exactly solvable \\
Russian doll BCS model of superconductivity
}
\author{Alberto Anfossi$^{1,3}$, Andr\'e  LeClair$^2$
 and  Germ\'an Sierra$^3$}
\affiliation{$^1$ Dipartimento di Fisica del Politecnico, Torino, Italy}
\affiliation{$^2$Newman Laboratory, Cornell University, Ithaca, NY}
\affiliation{$^3$Instituto
de F\'{\i}sica Te\'orica, UAM-CSIC, Madrid, Spain}

\date{February 2005}

\bigskip\bigskip\bigskip\bigskip

%
\font\numbers=cmss12
\font\upright=cmu10 scaled\magstep1
\def\stroke{\vrule height8pt width0.4pt depth-0.1pt}
\def\topfleck{\vrule height8pt width0.5pt depth-5.9pt}
\def\botfleck{\vrule height2pt width0.5pt depth0.1pt}
\def\Zmath{\vcenter{\hbox{\numbers\rlap{\rlap{Z}\kern
0.8pt\topfleck}\kern 2.2pt
                   \rlap Z\kern 6pt\botfleck\kern 1pt}}}
\def\Qmath{\vcenter{\hbox{\upright\rlap{\rlap{Q}\kern
                   3.8pt\stroke}\phantom{Q}}}}
\def\Nmath{\vcenter{\hbox{\upright\rlap{I}\kern 1.7pt N}}}
\def\Cmath{\vcenter{\hbox{\upright\rlap{\rlap{C}\kern
                   3.8pt\stroke}\phantom{C}}}}
\def\Rmath{\vcenter{\hbox{\upright\rlap{I}\kern 1.7pt R}}}
\def\Z{\ifmmode\Zmath\else$\Zmath$\fi}
\def\Q{\ifmmode\Qmath\else$\Qmath$\fi}
\def\N{\ifmmode\Nmath\else$\Nmath$\fi}
\def\C{\ifmmode\Cmath\else$\Cmath$\fi}
\def\R{\ifmmode\Rmath\else$\Rmath$\fi}

\begin{abstract}
The recently proposed Russian doll BCS model provides a simple example
of a many body system whose renormalization group analysis reveals
the existence of limit cycles in the running coupling constants of the
model. The model was first studied using RG,
mean field and numerical methods
showing the Russian doll
scaling of the spectrum,
$E_n \sim E_0 \;  e^{- \lambda n}$, where $\lambda$ is the RG period.
In this paper
we use the recently discovered exact solution of this model to study
the low energy spectrum.  We find  that, in addition
to the standard quasiparticles, the electrons can
bind into  Cooper pairs that are different from those
forming the condensate and with higher energy.
These excited Cooper pairs can be described by a
quantum number $Q$ which appears in the Bethe ansatz equation
and has a RG interpretation.

\end{abstract}


\maketitle

\vskip 0.2cm

%
%
%
%
\def\oti{{\otimes}}
\def\lb{ \left[ }
\def\rb{ \right]  }
\def\tilde{\widetilde}
\def\bar{\overline}
\def\hat{\widehat}
\def\*{\star}
\def\[{\left[}
\def\]{\right]}
\def\({\left(}      \def\BL{\Bigr(}
\def\){\right)}     \def\BR{\Bigr)}
    \def\BBL{\lb}
    \def\BBR{\rb}
%
%
\def\zb{{\bar{z} }}
\def\zbar{{\bar{z} }}
\def\frac#1#2{{#1 \over #2}}
\def\inv#1{{1 \over #1}}
\def\half{{1 \over 2}}
\def\d{\partial}
\def\der#1{{\partial \over \partial #1}}
\def\dd#1#2{{\partial #1 \over \partial #2}}
\def\vev#1{\langle #1 \rangle}
\def\ket#1{ | #1 \rangle}
\def\rvac{\hbox{$\vert 0\rangle$}}
\def\lvac{\hbox{$\langle 0 \vert $}}
\def\2pi{\hbox{$2\pi i$}}
\def\e#1{{\rm e}^{^{\textstyle #1}}}
\def\grad#1{\,\nabla\!_{{#1}}\,}
\def\dsl{\raise.15ex\hbox{/}\kern-.57em\partial}
\def\Dsl{\,\raise.15ex\hbox{/}\mkern-.13.5mu D}
%
%
\def\ga{\gamma}     \def\Ga{\Gamma}
\def\be{\beta}
\def\al{\alpha}
\def\ep{\epsilon}
\def\vep{\varepsilon}
\def\dep{d}
\def\arc{{\rm Arctan}}
\def\la{\lambda}    \def\La{\Lambda}
\def\de{\delta}     \def\De{\Delta}
\def\om{\omega}     \def\Om{\Omega}
\def\sig{\sigma}    \def\Sig{\Sigma}
\def\vphi{\varphi}
%
%
\def\CA{{\cal A}}   \def\CB{{\cal B}}   \def\CC{{\cal C}}
\def\CD{{\cal D}}   \def\CE{{\cal E}}   \def\CF{{\cal F}}
\def\CG{{\cal G}}   \def\CH{{\cal H}}   \def\CI{{\cal J}}
\def\CJ{{\cal J}}   \def\CK{{\cal K}}   \def\CL{{\cal L}}
\def\CM{{\cal M}}   \def\CN{{\cal N}}   \def\CO{{\cal O}}
\def\CP{{\cal P}}   \def\CQ{{\cal Q}}   \def\CR{{\cal R}}
\def\CS{{\cal S}}   \def\CT{{\cal T}}   \def\CU{{\cal U}}
\def\CV{{\cal V}}   \def\CW{{\cal W}}   \def\CX{{\cal X}}
\def\CY{{\cal Y}}   \def\CZ{{\cal Z}}

\def\rvac{\hbox{$\vert 0\rangle$}}
\def\lvac{\hbox{$\langle 0 \vert $}}
\def\comm#1#2{ \BBL\ #1\ ,\ #2 \BBR }
\def\2pi{\hbox{$2\pi i$}}
\def\e#1{{\rm e}^{^{\textstyle #1}}}
\def\grad#1{\,\nabla\!_{{#1}}\,}
\def\dsl{\raise.15ex\hbox{/}\kern-.57em\partial}
\def\Dsl{\,\raise.15ex\hbox{/}\mkern-.13.5mu D}
%
%
%
\font\numbers=cmss12
\font\upright=cmu10 scaled\magstep1
\def\stroke{\vrule height8pt width0.4pt depth-0.1pt}
\def\topfleck{\vrule height8pt width0.5pt depth-5.9pt}
\def\botfleck{\vrule height2pt width0.5pt depth0.1pt}
\def\Zmath{\vcenter{\hbox{\numbers\rlap{\rlap{Z}\kern
0.8pt\topfleck}\kern 2.2pt
                   \rlap Z\kern 6pt\botfleck\kern 1pt}}}
\def\Qmath{\vcenter{\hbox{\upright\rlap{\rlap{Q}\kern
                   3.8pt\stroke}\phantom{Q}}}}
\def\Nmath{\vcenter{\hbox{\upright\rlap{I}\kern 1.7pt N}}}
\def\Cmath{\vcenter{\hbox{\upright\rlap{\rlap{C}\kern
                   3.8pt\stroke}\phantom{C}}}}
\def\Rmath{\vcenter{\hbox{\upright\rlap{I}\kern 1.7pt R}}}
\def\Z{\ifmmode\Zmath\else$\Zmath$\fi}
\def\Q{\ifmmode\Qmath\else$\Qmath$\fi}
\def\N{\ifmmode\Nmath\else$\Nmath$\fi}
\def\C{\ifmmode\Cmath\else$\Cmath$\fi}
\def\R{\ifmmode\Rmath\else$\Rmath$\fi}

\def\barray{\begin{eqnarray}}
\def\earray{\end{eqnarray}}
\def\beq{\begin{equation}}
\def\eeq{\end{equation}}

\def\no{\noindent}

\def\gpar{g_\parallel}
\def\gperp{g_\perp}

\def\Jb{\bar{J}}
\def\dx{\frac{d^2 x}{2\pi}}

\def\rap{\beta}
\def\s{\sigma}
\def\spec{\zeta}
\def\comb{\frac{\rap\theta}{2\pi} }
\def\Ga{\Gamma}

\def\L{{\cal L}}
\def\g{{\bf g}}
\def\K{{\cal K}}
\def\I{{\cal I}}
\def\M{{\cal M}}
\def\F{{\cal F}}

\def\gpar{g_\parallel}
\def\gperp{g_\perp}
\def\Jb{\bar{J}}
\def\dx{\frac{d^2 x}{2\pi}}
\def\imag{\Im {\it m}}
\def\real{\Re {\it e}}
\def\Jbar{{\bar{J}}}
\def\kh{{\hat{k}}}

\section{Introduction}

The existence of limit cycles in the renormalization group 
(RG)  is a possibility considered  
long ago by Wilson
in the framework of High Energy Physics~\cite{Kwilson}, 
but only recently  has it  found 
a concrete realization in several models in
nuclear physics \cite{nuclear},
quantum field theory \cite{BLflow,LRS},
quantum mechanics \cite{GW},
superconductivity \cite{RD},
S-matrix models \cite{log,ellipticS},
Bose-Einstein condensation \cite{Bose},
effective low energy  QCD \cite{QCD}, 
few body systems and Efimov states \cite{nuclear,few-body}, etc
(for a review of see \cite{few-body}.) 
The subject of duality cascades in supersymmetric
gauge theory \cite{Klebanov}
is also suggestive of limit-cycle behavior. 
Chaotic flows have also been recently
considered \cite{GW,morozov}.  The concepts of discrete
scale invariance and quantum groups with $q$ real
are also closely related \cite{log,Tierz}.

A RG limit cycle means that the coupling constants of the model
are invariant under  a finite RG transformation. There are
generically two types of RG limit cycles: infrared and ultraviolet.
In the former the limit cycles appear in the RG flow towards low
energy and imply peculiar scaling properties of the spectrum termed
as Russian doll scaling in \cite{RD} for obvious reasons. In
particular if there are bound states, their energies $E_n \;
(n=0,1,\dots)$ will scale as $e^{-n \lambda}$ where $\lambda$ is the
``time'' needed to complete an RG cycle. The ultraviolet RG limit
cycles appear in the RG flow towards high energy and lead to
log-periodic behavior of the scattering as a function of energy,
and/or 
 Russian doll behavior in  the masses of resonances
$M_n \sim e^{n \lambda}$ if they are present.

Reference \cite{RD} proposed a slight modification of the
BCS model of superconductivity, referred to here as the 
RD model,  
whose RG analysis revealed the existence of infrared
limit cycles. The standard BCS model is given by the sum of
a kinetic Hamiltonian describing the propagation of free electrons,
plus a pairing Hamiltonian describing
the scattering of
a pair of electrons occupying time reversed states,
say $({\bf k'}, \uparrow)$
and  $(-{\bf k'}, \downarrow)$, into another pair of states,
say  $({\bf k}, \uparrow)$ and  $(-{\bf k}, \downarrow)$,
with an amplitude $V_{{\bf k},{\bf k'}}$ \cite{BCS,BCS-book}.
For $s$-wave pairing one can approximate the matrix element
$V_{{\bf k},{\bf k'}}$ by a constant value $G$ for all
the incoming ${\bf k'}$
and outgoing momenta ${\bf k}$  within the Debye shell around
the Fermi surface. To perform an RG analysis one
defines a dimensionless BCS coupling constant
$ g = G N(0)$ with $N(0)$ the energy density at the Fermi level.
Then under the RG flow $g$ increases towards the IR becoming infinite
at a scale signaling the formation of Cooper pairs.
 The modification added in \cite{RD} was to let the
amplitude   $V_{{\bf k},{\bf k'}}$ to pick up an extra imaginary
piece proportional to $\pm i h$ where the sign depends on wether the
outgoing pair has greater or lower energy than the incoming pair
($h$ is a dimensionless coupling constant). This of course breaks
the time reversal symmetry which is a characteristic feature of the RD
model. It was shown in \cite{RD} that the coupling $h$ remains
invariant under the RG flow while $g$ exhibits a cyclic behavior
with an RG period given by $\lambda = \pi/h$. Under this flow the
coupling $g$ becomes infinite at a finite scale but its value can be
continued through minus infinity until it reaches its original value
after the RG period $\lambda$. One may expect from this behavior the
existence of infinitely many scales related by $\lambda$. Indeed,
using the mean field BCS ansatz, which is valid for generic choices
of the pair amplitudes
  $V_{{\bf k},{\bf k'}}$,  one can show
that the gap
equation admits infinitely many solutions characterized by the gaps
$\Delta_n = \Delta_0 \; e^{- n \lambda} \; (n=0,1,\dots)$, where
the ground state corresponds to
the solution with the highest value of the gap, $\Delta_0$,
and the other solutions, $\Delta_{n > 0}$, are high energy collective
excited states. Let us call the latter solutions
of the BCS gap equation the ``superconduting dolls'' or simply
``dolls'' and the integer $n$
that label them the ``nesting number''. The larger the nesting number,
the larger is the size of the Cooper pairs forming the collective state,
which is given by the correlation length
$\xi_n = \xi_0 \;  e^{ n \lambda}$.

These results showed  in a simple case the intimate relationship
between the cyclic properties of the RG flows and the spectrum of
the theory, which were also confirmed numerically in the case of one
Cooper pair~\cite{RD}. The case of more pairs was more difficult to
deal with numerically.
The DMRG or other ground state numerical methods were not helpful 
since one needs to explore high energy excited states to
identify the ``dolls''. Fortunately  the Russian doll BCS
model has been recently shown to be exactly solvable by Dunning and
Links using the Quantum Inverse Scattering Method \cite{links}. This
important result will enable us to explore in  detail the
spectrum of the RD model by solving the Bethe ansatz equations
(BAE). This is the aim of this paper, whose organization is as
follows.

In section II we generalize the definition of the
RD model, as given in  \cite{RD}, showing that
the manifold of solutions of the gap equation with
Russian doll scaling is a generic feature.
In section III we review the Dunning and Links exact solution
and show its agreement in the large $N$ limit
with the mean field solution obtained in section II
($N$ is the number of particles).
In section IV we focus on the solution of the BAE for the one
Cooper pair problem,  which serves to clarify some conceptual
and technical issues before considering the many body case.
In section V we explore numerically and analytically
the low energy spectrum of the model in the large
$N$ limit, finding  new types
of elementary excitations which are absent in the standard BCS model.
These  new kinds  of elementary excitations can be characterized by
a quantum number  $Q$  which has a clear  meaning in both the 
RG and in the exact Bethe ansatz solution.
In appendix A we review the 
analytic techniques
used in section V and in appendix
B we describe the elementary excitations of the standard
BCS model in the canonical ensemble to facilitate the comparison
with the excitations found in the RD model.

\section{The Russian doll BCS model: Mean Field Solution}

The model defined in  \cite{RD}  was based on a modification of the
BCS model used to described the ultrasmall superconducting grains
discovered by Ralph, Black and Tinkham \cite{RBT} (see
\cite{vDR} for a review). 
The latter model, also known as the picket-fence model
in Nuclear Physics, has doubly degenerate discrete electronic energy
levels which can be taken equally spaced or randomly distributed. It
was claimed in \cite{RD} that the results obtained for the equally
spaced model are valid in more general circumstances. We shall show
in this section that this is indeed the case by considering a
version of the RD model that is more similar  to the standard
formulation of the BCS model.

Let us call $c_{k, \sigma}$ ($c_{k, \sigma}^\dagger$)
the destruction (creation) operator of an electron
in the state with momenta ${\bf k}$ and spin
$\sigma =  \uparrow, \downarrow$.
The BCS pairing Hamiltonian (also called reduced Hamiltonian)
is given by \cite{BCS-book}
\beq
H = \sum_{k, \sigma} \ep_k \; n_{k,\sigma} +
\sum_{k,k'} V_{k,k'}  \; b^\dagger_{k} \; b_{k'}, 
\label{a1}
 \eeq
\no
where $\ep_k$ is the energy
of an electron with momenta $\bf{k}$,
$n_{k,\sigma} = c^\dagger_{k ,\sigma} c_{k,\s}$
is the number operator of the state $(\bf{k} , \s)$,  and
$b^\dagger_{k}$ and $b_k$ are the creation and
destruction operators of a pair of electrons
in the states $(\bf{k}, \uparrow)$ and $(- \bf{k}, \downarrow)$
\beq
b_k^\dagger = c^\dagger_{k, \uparrow} c^\dagger_{-k, \downarrow}, \qquad
b_k = c_{-k, \downarrow} c_{k, \uparrow}. 
\label{a2}
\eeq
The BCS ansatz for the  ground state in the grand
canonical ensemble is given by
\beq
|\psi_0 \rangle = \prod_k (u_k + v_k\,  b^\dagger_k) |0 \rangle, 
\label{a3}
\eeq
\no where $|0 \rangle$ is the Fock vacuum of the
fermion operators and $u_k, v_k$ are variational parameters
whose values are given by,
\barray
 u_k^2 &  = &  \frac{1}{2} \left( 1 + \frac{ \xi_k}{E_k} \right),
\qquad
v_k^2   =   \frac{1}{2} e^{ 2 i \phi_k}
\left( 1 - \frac{ \xi_k}{E_k} \right), \nonumber \\
E_k & = & \sqrt{ \xi_k^2 + |\Delta_k|^2}, \qquad
\xi_k = \ep_k - \mu + V_{k,k}. 
\label{a4}
\earray
\no  The gap function $\Delta_k$ and the chemical
potential $\mu$ are found by solving 
the gap equation
\beq
\Delta_k = - \sum_{k'} V_{k,k'} \; \frac{\Delta_{k'}}{2 E_{k'}},
\qquad \Delta_k \equiv  |\Delta_k| e^{ i\phi_k}, 
\label{a5}
\eeq

\no and the chemical potential equation,
\beq
N_e = 2 \sum_k   |v_k|^2 = \sum_k \left( 1 - \frac{ \xi_k}{E_k} \right), 
\label{a6}
\eeq

\no where $N_e$ is the number of electrons. If $V_{k,k'}$ is real,
so is the gap function $\Delta_k$, up to an overall phase factor
which can be chosen equal to 1 (i.e. $\phi_k =0$). In the cases
where the system possesses particle-hole symmetry around the Fermi
surface the chemical potential coincides with the Fermi energy
$\ep_F$. Measuring all energies relative to $\ep_F$ one can take
$\mu =0$ \cite{BCS-book}. We shall suppose that this is the case.

The  Russian doll BCS model is defined in terms of the scattering
potential:
\beq
\label{a7}
- V_{k, k'} = \left\{
\begin{array}{lc}
G + i \eta \;  {\rm sign}(\ep_k - \ep_{k'})
& \quad |\ep_k|
\; {\rm and} \; |\ep_{k'}| < \omega_c \\
0            & {\rm otherwise} \\
\end{array}
\right.
\eeq

\no where ${\rm sign}(x)$ is the sign function. This potential
describes for $G > 0$ the attraction of electrons within a shell of
width $2 \omega_c$ centered around the Fermi surface plus an
imaginary term which depends on the sign of the difference between
the energies of the incoming and outgoing electrons. Since
$V^*_{k,k'} = V_{k',k}$ the Hamiltonian is hermitian but the
imaginary term breaks the time reversal symmetry.
Setting $\eta =0$ we recover the standard model used
to describe $s-$wave superconductors \cite{BCS-book}.
Obviously the $\eta$
interaction can be generalized to other type of symmetries such as
$p$-wave, $d$-wave, etc. Here we shall only consider  the $s$-wave case.

Let us next solve the gap equation (\ref{a5}) for
a large number of electrons where the discrete energy
levels become a continuum. If we denote by $N(\ep)$ the density of levels
per energy then eq.(\ref{a5}), for the potential (\ref{a7}), becomes
\beq
\Delta(\ep) = \int_{- \omega_c}^{\omega_c}
d \ep' \; N(\ep') \[ G + i  \eta \; {\rm sign} (\ep - \ep')
\] \; \frac{ \Delta(\ep')}{2 E(\ep')}
\label{a8}
\eeq
\no  where $E(\ep) = \sqrt{ \ep^2 + |\Delta(\ep)|^2}$
if we assume particle-hole symmetry ($\mu = \ep_F= 0$).
Notice that $\Delta_k \equiv \Delta(\ep)$
depends on the momenta $k$ through its energy $\ep = \ep(k)$.
In the standard BCS model where $\eta =0$,
eq.(\ref{a8}) implies a constant gap, i.e. $\Delta(\ep) = \Delta_0$,
whose value is given by the solution of the integral,
\beq
\frac{1}{G} = \int_0^{\omega_c} d \ep \; \frac{N(\ep)}{
\sqrt{ \ep^2 + \Delta_0^2}}
\label{a9}
\eeq
\no where we have used again particle-hole symmetry namely,
i.e. $N(\ep) = N(- \ep)$.
Approximating the density $N(\ep)$ by its value at the
Fermi level, $N(0)$,  one can perform the integral (\ref{a9})
obtaining the well known result:
\beq
\Delta_0 = \frac{\omega_c}{\sinh( 1/G N(0))} \sim 2 \omega_c
e^{-{1}/{G N(0)}}
\label{a10}
\eeq
The last expression  is valid in
the weak coupling case $G N(0) < 1/4$.

When $\eta \neq 0$ the gap $\Delta(\ep)$ depends
on the energy. Differentiating  eq.(\ref{a8}) with
respect to $\ep$ we find,
\beq
\frac{ d \Delta(\ep)}{d \ep} = i \eta \;  \frac{ N(\ep) \Delta(\ep)}{
E(\ep)}
\label{a11}
\eeq

\no Hence the modulus of $\Delta(\ep)$ remains constant, $\Delta$,
and  its phase varies with $\ep$, i.e.
\beq
\Delta(\ep) = \Delta \; e^{i \phi(\ep)}, \qquad
\frac{ d \phi(\ep)}{d \ep} = \eta \;  \frac{ N(\ep)}{
E(\ep)}
\label{a12}
\eeq

\no Without loss of generality we can choose $\phi(0) =0$ so that
\beq
\phi(\ep) = \eta  \int_0^\ep d \ep' \, \frac{ N(\ep')}{E(\ep')}
\label{a13}
\eeq

The value of $\Delta$ is found by imposing
eq.(\ref{a8}) at the Fermi energy, $\ep=0$. Using eq.
(\ref{a12}) one can trade the integral over $\ep$ into an integral
over the phase $\phi$,
\beq
1 = \int_{-\phi_c}^{\phi_c} \frac{d \phi}{2 \eta}
\;  (G - i \eta
\;   {\rm sign}(\phi)) \; e^{ i \phi}
= 1 - \cos \phi_c + \frac{G}{\eta} \, \sin \phi_c
\label{a14}
\eeq

\no
where $\phi_c \equiv \phi(\omega_c)$ is the value of the phase
of the order parameter at the boundary of the Debye shell.
The equation satisfied by $\phi_c$ is then
\beq
\tan \phi_c = \frac{\eta}{G} \Longrightarrow \phi_c =
{\rm Arctan} \left( \frac{\eta}{G} \right) + \pi Q
\label{a15}
\eeq

\no where {\rm Arctan} is the principal determination
and $Q$ is an integer. Plugging (\ref{a15})  into (\ref{a13})
(for $\ep = \omega_c$) one gets,
\beq
 \frac{1}{\eta} \[{\rm Arctan} \left( \frac{\eta}{G} \right) + \pi Q \]
= 
\int_0^{\omega_c} d \ep \, \frac{ N(\ep)}{
 \sqrt{ \ep^2 + \Delta_{Q}^2}}
\label{a16}
\eeq
\no which is the equation that fixes 
$\Delta \equiv  \Delta_Q$ as a
function of $Q$, $G$ and $\eta$. The positiveness of the RHS of this
equation implies that $Q$ must be a non negative integer. For each
value of $Q=0, 1 ,\dots$ eq.(\ref{a16})  is identical to eq.
(\ref{a9}) with an effective value of the coupling constant $G_Q$
given by
\beq
\frac{1}{G_{Q}} =
 \frac{1}{\eta} \[{\rm Arctan} \left( \frac{\eta}{G} \right) + \pi Q \],
\qquad Q=0,1, \dots
\label{a17}
\eeq
\no
Hence in the weak coupling regime the gaps
$\Delta_Q$ satisfy the Russian doll scaling
\beq
\Delta_{Q} \sim 2 \omega_c e^{- 1/G_Q N(0)} \sim
\Delta_0 e^{- \lambda Q}
\label{a18}
\eeq
where
\beq
\lambda =
\frac{\pi}{h}, \;\; \qquad  h = \eta N(0)
\label{a18bis}
\eeq
can be identified with the ``time'' it takes the RG to complete
a cycle \cite{RD}.
In the limit $\eta \rightarrow 0$ we get that $G_{Q=0} \rightarrow G$
and the solution $Q=0$ coincides with the unique BCS solution
while the states with $Q > 0$ have a vanishing gap and 
approach the asymptotically the Fermi state.

\begin{figure}[t!]
\begin{center}
\includegraphics[height= 3 cm,angle= 0]{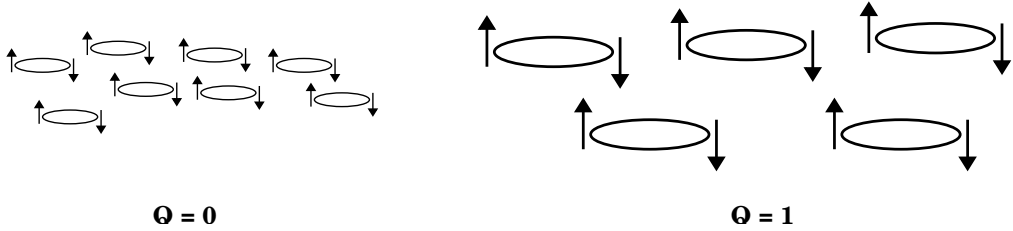}
\end{center}
\caption{Pictorial representation of two solutions of the
RD equations with $Q=0$ (ground state) and $Q=1$ (high energy state).
The size of the Cooper pairs for $Q=0$, $\xi_0$, is smaller
that the size of the $Q=1$ pairs, $\xi_1$, since eq.(\ref{a18})
implies $\xi_1 = e^\lambda \xi_0$.
}
\label{condensates}
\end{figure}

The ground state of the RD model is the solution
with the lowest total energy or equivalently  with the lowest
condensation energy
$E_c$.
The latter quantity is defined as the difference between the
energy of the ground state and the energy of the Fermi state,
 and it is given
for the BCS state in the weak coupling regime  
by $E_c= - \frac{1}{2} N(0) \Delta^2$.
Hence the ground state corresponds to the $Q=0$ solution, while the
states with $Q > 0$ are high energy excited states
(see fig. \ref{condensates}).

The quantum number $Q$ not only determines the modulus of the
superconducting order parameter $\Delta_Q$ but also its phase.
Eq.(\ref{a15}) implies that  the phase variation
of $\phi(\ep)$ from the bottom to the top of the Debye shell
is given by $2 \pi Q$, up to a constant,
so that $Q$ is a sort of winding number.
This interpretation will be confirmed by the exact solution.
These results generalize those obtained in \cite{RD} to arbitrary
energy densities $N(\ep)$ and  therefore are
valid in two and three dimensions. In the 1D case considered
in  \cite{RD}  $N(0)$ is given by $1/d$, hence
$g = G N(0) =G/d$ and $h = \eta N(0) = \eta/d$,
are dimensionless couplings. In more general cases where
the matrix element $V_{k,k'}$ depends on the momenta
we also expect a RD behaviour as long as time reversal
is broken, $V_{k,k'}^* = V_{k',k} $, the reason being 
that the Physics is dominated by the vicinity of the Fermi
surface in which case the simplified model (\ref{a7})
is a good approximation. This points towards the universality
of the RD model, whose experimental realization 
seems a priori feasible.

\section{Relation between the exact
and mean field solutions}

We shall next summarize the exact solution of the Russian doll BCS
model obtained by Dunning and Links \cite{links}. We will
use  the notation  adapted to the study of ultrasmall
superconducting grains,  although the results are more general as we
have shown  in the previous section. For grains,  the single
particle basis of momenta and spin is not appropriate due to the
lack of translational invariance. However one can still use time
reversed states $|j, \pm \rangle$ created and destroyed by  fermion
operators $c^\dagger_{j,\pm}$ and $c_{j,\pm}$ where $j=1, \dots, N$
labels $N$ discrete energy levels $\vep_j$.
 The energy $\vep_j$  represents the energy of
a pair of electrons in
a given level (it is
twice the single particle energy $\ep_k$ of the previous section).
In the equally spaced model
the distance between two consecutive levels is fixed to
$2 \dep$, i.e. $\vep_{j+1} - \vep_j = 2 \dep$, so that
$-\omega < \vep_j < \omega$ where
$\omega = N \dep$ is twice the Debye energy $\omega_c$.
Henceforth all the energies  will be
twice their standard values.

Let $b_j = c_{j,-} c_{j,+}$,
 $b_j^\dagger  = c_{j,+}^\dagger  c_{j,-}^\dagger$
denote the usual pair operators.
Using this  notation  the Hamiltonian (\ref{a1})
becomes
\beq
\label{b1}
H =  \sum_{j=1}^N \vep_j ~ b^\dagger_j b_j ~ + ~
\sum_{j,j' = 1}^N ~
V_{jj'} \> b^\dagger_j b_{j'} ,
\eeq
where $V_{jj'}$ is the scattering potential. Similarly
to eq.(\ref{a7}),  the RD model is defined by:
\beq
\label{b2}
- V_{jj'} = G + i \, \eta \, {\rm sign}  (\vep_j - \vep_{j'}). 
\eeq

\no In (\ref{b2}) we are assuming that all the
pair energies $\vep_j$ are
different and labeled in increasing order, i.e.
$ \vep_j < \vep_{j'}$ if $ j < j'$. Hence the second term in
(\ref{b2}) is equivalent to ${\rm sign}(j - j')$ so that
(\ref{b1}) can be written as
\beq
\label{b3}
H =  \sum_{j=1}^N (\vep_j - G)  ~ b^\dagger_j b_j ~ - ~
\bar{G} \sum_{j  > k = 1}^N ~ ( e^{ i \alpha}  b^\dagger_j b_{k}
+ e^{ - i \alpha}  b^\dagger_k b_{j} )
\eeq

\no where
\beq
\alpha = \arc \left( \frac{\eta}{G} \right),
\qquad \bar{G} = \sqrt{G^2 + \eta^2}
\label{b4}
\eeq

\no The hamiltonian (\ref{b3}) is basically the one
studied  in \cite{links}, where it  was shown to be
exactly solvable using the Quantum Inverse Scattering Method (QISM).
These authors showed that (\ref{b3})
appears in the transfer matrix of a certain vertex model
as the second order term in a series expansion
in the inverse of the spectral parameter.

The standard BCS model, which corresponds to the limit
$\eta \rightarrow 0$ of the RD model, is also exactly
solvable and integrable. Its exact solution was obtained long ago
by Richardson \cite{R1,RS,R-roots}
and the integrals of motion found much later
by Cambiaggio, Rivas and Sarraceno \cite{CRS}
(for a review see \cite{RMP}).
These results were rederived
in the framework of the
QISM in terms of an inhomogeneous XXX vertex model with twisted
boundary conditions \cite{AFF,ZLMG,vDP,Scomo}.
In this approach
the hamiltonian and the  conserved quantities appear in
a semiclassical expansion of the transfer matrix in
a parameter, $\eta' \rightarrow 0$, that enters in the XXX vertex
$R(u)$-matrix defining the model. Expanding the $R(u)$-matrix in
this parameter, i.e. $R(u) = 1 + \eta' \; r(u) + O(\eta'^2)$,
yields the classical $r(u)$-matrix which satisfies
the classical Yang-Baxter equation,
while the $R$-matrix
satisfies the quantum Yang-Baxter equation.
As we shall see below the
latter parameter $\eta'$
can be identified with the parameter $\eta$
of the Russian doll BCS model.
Hence,  from the viewpoint of quantum integrability,  the RD model
is the  ``quantum'' version of the standard 
BCS model which arises in the semiclassical 
limit:
\beq
\lim_{\eta \rightarrow 0} \; {\rm RD} \; {\rm model} 
= {\rm BCS} \; {\rm model} 
\label{RD}
\eeq

Let us summarize the exact solution of the hamiltonian
(\ref{b1},\ref{b3}) obtained in  \cite{links}.
The eigenstates in the sector with $M$ electron
pairs are found by solving the BAE,
\beq
e^{2 i \alpha}
\prod_{j=1}^N \frac{E_a - \vep_j + i \eta}{E_a - \vep_j - i \eta}
= \prod_{b=1 (\neq a)}^M
 \frac{E_a - E_b + 2 i \eta}{E_a - E_b - 2 i \eta}
\label{b5}
\eeq

\no where the ``rapidities''  $E_a \; (a=1, \dots, M)$
give the total energy of the state as,
\beq
E = \sum_{a=1}^M \; E_a
\label{b6}
\eeq

\no In the limit where $\eta \rightarrow 0$ we see from
(\ref{b4}) that $\alpha \sim \eta/G$ and then (\ref{b5})
becomes
\beq
\frac{1}{G} + \sum_{j=1}^N \frac{1}{E_a - \vep_j}
- \sum_{b=1 (\neq a)}^M \frac{2}{ E_a - E_b}
= 0
\label{b7}
\eeq
These  are the Richardson equations whose solution give
the eigenstates of the BCS Hamiltonian
(i.e. $\eta =0$).
To solve the BAE (\ref{b5}) one can proceed as in the study
of the antiferromagnetic Heisenberg spin 1/2 chain by
first taking the logarithm of the equations. Choosing the branch
of the logarithm,
\beq
\frac{1}{2 i} \log \frac{ x + i \eta}{x - i \eta} =
\arc\left( \frac{\eta}{x} \right)
\label{b8}
\eeq

\no
we obtain from (\ref{b5})
\beq
\alpha + \pi Q_a  + \sum_{j=1}^N \arc \left(
 \frac{\eta}{E_a - \vep_j} \right)
- \sum_{b=1 (\neq a)}^M \arc \left(
\frac{2 \eta}{ E_a - E_b} \right)
= 0
\label{b9}
\eeq

\no where $Q_a \;  (a=1, \dots, M)$ are a set of integers.
In the case of spin chains there is also a set of integers
$Q_j$ associated to the rapidity variables whose choice
determines the ground
state and the excitations. For example the ground state
of the antiferromagnetic spin chains
corresponds to increasing $Q's$ ($Q_j = j + const.$), while
holes in that distribution correspond to the elementary
excitations of the model, i.e. the spinons \cite{Faddeev}.

Let us first show that the mean field solution of the
Russian doll BCS obtained
in section II coincides with the exact solution
to leading order in $N$.
We first need to recall
that the mean field solution of the
BCS model agrees to leading order in $N$ 
with Richardson's exact solution.
This was proved by Gaudin \cite{G-book}
and Richardson \cite{R-limit} who solved equation (\ref{b7})
in the large $N$ limit using an electrostatic analogy
(see \cite{largeN}
for the comparison between the analytical
and numerical solutions, which is reviewed in 
the Appendix A).
In the Gaudin and Richardson approaches the eq. (\ref{b7})
yield the BCS gap equation (\ref{a9}).
The thermodynamic limit amounts to letting the number of levels
$N$ go to infinity and the energy spacing $d$ go
to zero while $\omega = N d$ is kept constant.
Eq.(\ref{b7}) implies that $E_a \sim N G \sim N d = \omega$. 
On the other hand $\eta \sim d \sim 1/N$. 
Hence to leading order in $N$ eq. (\ref{b9}) 
becomes, 
\beq
\frac{1}{\eta}( \alpha + \pi Q_a) + \sum_{j=1}^N \frac{1}{E_a - \vep_j}
- \sum_{b=1 (\neq a)}^M \frac{2}{ E_a - E_b}
= 0
\label{b10}
\eeq
The next to leading corrections in $1/N$ of this
formula can be computed
using the results of \cite{strings}, but shall not considered
here. Taking $Q_a = Q \;$ for $a=1, \dots, M$,
then eq.(\ref{b10}) becomes the Richardson eq.(\ref{b7})
with a $Q$ dependent value of the coupling constant
\beq
\frac{1}{G_{Q}} = \frac{1}{\eta} \left( \alpha + \pi Q \right)
=  \frac{1}{\eta} \left[ \arc \left( \frac{\eta}{G} \right)
 + \pi Q \right]
\label{b11}
\eeq

\no where we have used eq.(\ref{b4}).
Hence the exact solution,  to leading order
in $N$, is given by the mean field solution with an effective value
of the BCS coupling constant given by eq.(\ref{b11}). On the other
hand this equation coincides with (\ref{a17}), which shows that
the mean field parameter $Q$, which labels the solutions
of the gap equation, can be identified with
the Bethe numbers $Q_a= Q$, common to all the roots $E_a$,
which classifies the solutions of the BAE.
This explains the origin of the quantum number $Q$
in the exact solution,  but it also suggests another possibility,
namely the existence of states where $Q_a$
depends on the root $E_a$.  If we are interested in 
the low energy spectrum of the ground state it is clear
that we must focus  on the cases where
$Q_a =0$ for almost all values of $a$ 
except  for a finite and small number.

\section{Exact solution of the one Cooper pair problem}

As for the usual BCS model, it is useful to first consider
the one Cooper pair problem, which is what we do in this section.
Remarkably,  the Bethe ansatz equations can be derived from
the RG.  

\subsection{RG derivation of the BAE}

The BAE (\ref{b9}) for a single Cooper pair (i.e. $M=1$)
with energy $E$ reduces to:
\beq
 \alpha  +   \pi Q +
 \sum_{j=1}^N \arc \left(
 \frac{\eta}{E - \vep_j} \right)
= 0
\label{rg1}
\eeq
\no
The  solutions are the eigenvalues $E$
of the Schr\"odinger equation
\beq
(\vep_j - G - E) \;
\psi_j = ( G + i \eta) \; \sum_{\ell =1}^{j-1}
\psi_\ell +  ( G - i \eta) \;  \sum^{N}_{\ell= j+1}
\psi_\ell, \qquad j=1, \dots, N
\label{rg2}
\eeq
\no
where $\psi_j$ is the wave function of the one-pair
state, i.e. $\sum_{j=1}^N \psi_{j}  \, b^\dagger_j \, | 0 \rangle$.
It is possible to  derive eq.(\ref{rg1})
using the RG method of Glazek and Wilson, which consists in the 
Gauss elimination 
of degrees of freedom in a quantum mechanical problem \cite{GW}.
Using eq.(\ref{rg2}) one first
eliminates the high energy component $\psi_{N}$ in terms
of the low energy ones,
\beq
\psi_N = \frac{G + i \eta}{\vep_N - G - E} \;
\sum_{\ell=1}^{N-1} \psi_\ell
\label{rg3}
\eeq
Replacing this eq. back
into eq.(\ref{rg2}) yields an eq. for the
remaining components,
\beq
(\vep_j - G_{N-1} - E) \;
\psi_j = ( G_{N-1} + i \eta) \; \sum_{\ell =1}^{j-1}
\psi_\ell +  ( G_{N-1} - i \eta) \;  \sum^{N-1}_{\ell= j+1}
\psi_\ell, \qquad j=1, \dots, N-1
\label{rg4}
\eeq
\no
where $G_{N-1}$ is related to $G$ and $\eta$ by the
eq.
\beq
G_{N-1} =  G_N + \frac{ G_N^2 + \eta^2}{\vep_N - G_N - E},
\qquad G_N = G
\label{rg5}
\eeq
Notice that $\eta$ remains invariant under
the discrete RG transformation.
Defining the quantities
\beq
\tan \alpha_n = \frac{\eta}{G_n}, \qquad
\tan \beta_j = \frac{\eta}{E - \vep_j}
\label{rg6}
\eeq
\no
eq.(\ref{rg5}) becomes,
\beq
\tan \alpha_{N-1} = \tan( \alpha_N + \beta_N)
\Rightarrow
\alpha_{N-1} = \alpha_N + \beta_N  \;\;  ({\rm mod} \;  \pi)
\label{rg7}
\eeq
\no
where $\alpha_n$ is  the principal
part of the ${\rm arctan}$ ($\alpha_n = {\rm Arctan}(\eta/G_n)$), etc.
After $p$-RG steps one gets,
\beq
\alpha_{N-p} = \alpha_N + \sum_{j=0}^{p-1} \beta_{N-j} +  \pi Q_p,
\qquad 0 \leq p < N-1
\label{rg8}
\eeq

\no
where $Q_p$ is an integer.
In the $p=N-1$ step the Sch\"odinger eq.(\ref{rg4}) reduces
to $(\vep_1 - G_1 - E) \psi_1 =0$.
The component $\psi_1$ cannot be zero since otherwise
the wave function  would vanish. This implies
that  $(\vep_1 - G_1 - E)=0$. Using eqs.
(\ref{rg5}) and (\ref{rg6}) the latter condition
is formally equivalent to the eqs. $G_0=\infty$ and $\alpha_0 = 0$.
With the latter formal definitions eq.(\ref{rg8})
can be extended to $p=N$,
\beq
0 = \alpha_N  + \sum_{j=1}^{N} \beta_{j} + \pi Q_N
\label{rg9}
\eeq

\no
which coincides with the BAE
(\ref{rg1}) upon  the identifications
$\alpha_N = \alpha$ and $Q_N= Q$.

The term $\pi Q$ in eq. (\ref{rg9}) now 
has a RG interpretation.  
For low energy bound states, i.e.
$E = \vep_0 \lesssim  \vep_1$,  eq. (\ref{rg8}) becomes
\beq
{\rm Arctan} \left(
\frac{\eta}{G_{N-p}} \right) =
{\rm Arctan} \left(
\frac{\eta}{G_N} \right) -
\sum_{j=0}^{p-1}
{\rm Arctan} \left(
\frac{\eta}{\vep_{N-j} - \vep_0} \right) + \pi Q_p
\label{rg10}
\eeq
\no
which can be viewed as a RG  equation for the couplings.
A cyclic RG flow is obtained if $G_{N-p} = G_N$, which happens
under the condition,
\beq
\pi Q_p =
\sum_{j=0}^{p-1}
{\rm Arctan} \left(
\frac{\eta}{\vep_{N-j} - \vep_0} \right)
\label{rg11}
\eeq
\no
$Q_p$ can be identified with the number of RG cycles
after $p$ RG steps.
The total number of RG cycles upon integration
of all the degrees of freedom is given by
\beq
n_C^{(1)} \equiv Q_N =  \left[
\frac{1}{\pi} \sum_{j=1}^{N}
 {\rm Arctan} \left(
\frac{\eta}{\vep_{j}- \vep_0} \right) \right]
\label{rg12}
\eeq
\no where $[ \dots]$ stands for the integer part. For the
equally spaced model we shall take
$\vep_j = d ( 2 j - N - 1) \;\;(j=0, 1, \dots,N)$.
Hence in the large $N$ limit eq.(\ref{rg12}) yields,
\beq
n_C^{(1)} \sim \frac{h}{2 \pi} \log N, \qquad \;\; h\equiv \eta/d , ~~~~~~
N \gg 1
\label{rg13}
\eeq
This equation was derived
in  \cite{RD} by taking the continuum limit of the discrete
RG eq. (\ref{rg5}) which reads,
\beq
\frac{d g}{d s} = \frac{1}{2} (g^2 + h^2), ~~~~~g = G/d 
\label{rg14}
\eeq
\no
where the RG scale $s$ is given by $s = \log(N_0/N)$
and $N_0$ is the initial size of the system. The solution of this
eq.
\beq
g(s) = {h} \tan \[
\frac{h s}{2} + \tan^{-1} \left(\frac{g_0}{h} \right) \],
\qquad g_0 = g(0)
 \label{rg15}
\eeq

\no
shows that the effective coupling $g(s)$ is a periodic
function of the scale $s$ with a period given by
$\lambda_1 = 2 \pi/h$.
In the many body case the period of the RG is a half of the one
pair result, namely $\lambda = \pi/h$ and the eq.(\ref{rg13})
has to be replaced by \cite{RD}
\beq
n_C \sim \frac{h}{\pi} \log N, \qquad \;\; N >> 1
\label{rg16}
\eeq
Eqs. (\ref{rg13}) and  (\ref{rg16}) admit a very simple
derivation. First notice that the size of the system
after a RG cycle is given by $e^{- \lambda} N$. Hence
in $n_C$ cycles, with $1 \simeq e^{- \lambda n_C} N$, 
the system is reduced to a site. But in each cycle 
a bound state, in the one Cooper pair problem,  
or a ``doll'', in the many body problem (i.e. solution of the gap eq.),
is produced and thus $n_C$ must give
the number of bound states or ``dolls''. Eq.(\ref{rg16})
also implies that in order to have at least one ``doll'',
i.e. $n_C \geq 1$, the coupling $h$ must be bigger
than a critical value which depends on the size, 
$h_c = \pi/\log N$. If $N$ is for example the Avogadro
number then $h_{c} = 0.057$, which means that 
macroscopic samples may give rise to ``dolls'' even for
small values of the $h$-coupling.

\subsection{Numerical solution}
\begin{figure}[h!]
\begin{center}
\includegraphics[width= 11 cm,angle= 0]{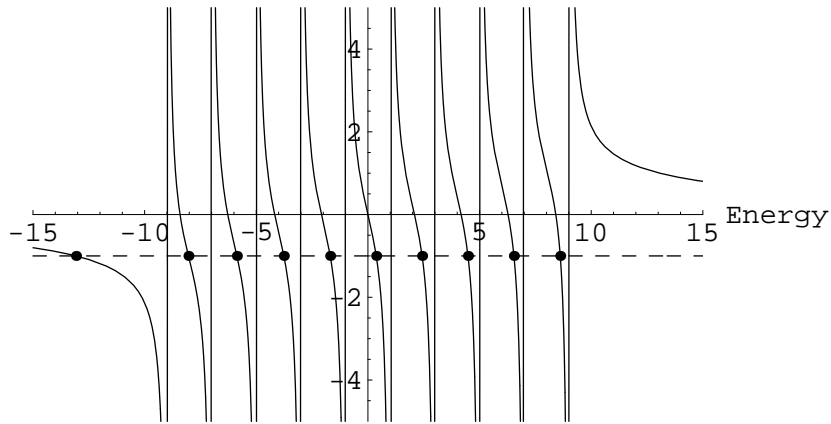}
\end{center}
\caption{Graphical solution of the eq.(\ref{rg17}) for a equally
spaced model for a coupling $G=1$ and $N=10$ levels. 
\cite{BCS-book}. The energies $E$ are given by
the abscisses of the points $\bullet$ which are the intersection of the
horizontal dotted line $-1/G$ with the function $ \sum_{j=1}^N
{1}/(E - \vep_j)$ with  $\vep_j = 2
j - N - 1 \;(d=1)$. For $G >0$ the lowest value of $E$ is 
always below
$\vep_1$ (Cooper pair state).} \label{tangents}
\end{figure}

The BAE (\ref{rg1}) in the limit $\eta \rightarrow 0$
becomes
\beq
- \frac{1}{G} = \sum_{j=1}^N \frac{1}{E - \vep_j}
\label{rg17}
\eeq
where we have choosen $Q=0$ in order to get a finite expression.
Eq.(\ref{rg17})  is the well known Cooper equation for a single pair
and it has $N$ solutions giving all the eigenstates
of the one pair Hamiltonian (see figure \ref{tangents}).

\begin{figure}[t!]
\begin{center}
\includegraphics[width= 12.5 cm,angle= 0]{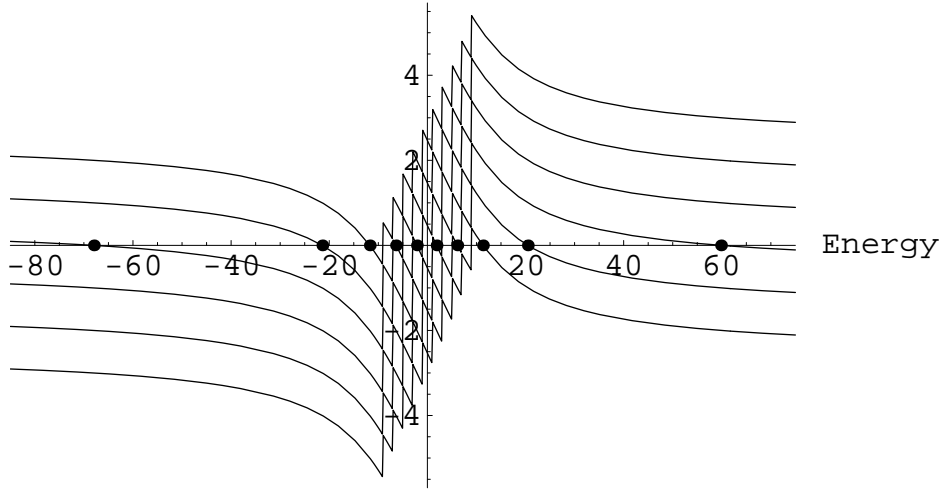}
\end{center}
\caption{Graphical solution of eq. (\ref{rg1})
for an equally spaced model
with $N=10$ energy levels 
($\vep_j = 2 j - N - 1, \;d=1)$ and 
couplings
$g= G=1, h= \eta =10$.
Every curve plots the LHS of eq.(\ref{rg1}) for
the values $Q=-3,-2,-1,0,1,2$ from bottom to top.
The solutions are the intersection of these curves
with the real axis (points denoted by $\bullet$).}
\label{onepair}
\end{figure}
Coming back to the RD model, eq.(\ref{rg1}), we show
in fig.~\ref{onepair} all the solutions
for a particular choice of parameters.
Every solution, $E_i$,  is characterized
by a given value of $Q_i$. Table 1
collects these values for
several examples. For small values of $h$
all the $Q's$ are zero as in the usual
Cooper problem (not displayed in table 1).
For $h=1$ there is  a single bound state with $Q=0$
and two high energy
states with $Q=-1$. For $h=3$ and 5  there are two bound
states with $Q=0$ and 1. Finally for $h=10$
there are three bound states with $Q=0,1,2$
(case depicted in fig.\ref{onepair}).
As shown in table 1, the number of bound states
with positive  value of $Q$
is equal to the number of RG cycles (\ref{rg12}).
This fact was explained above using limit cycle
RG arguments.
\begin{center}
\begin{tabular}{|c|c||c|c|c|c|c|c|c|c|c|c|}
  \hline
$h$ & $n_C^{(1)}$ & $Q_1$ & $Q_2$  & $Q_3$ & $Q_4$ & $Q_5$ &
 $Q_6$ & $Q_7$ & $Q_8$ & $Q_9$ & $Q_{10}$ \\ \hline
1 & 0 & $0^*$ & 0 & 0 & 0 & 0 & 0 & 0 & 0 & -1 & -1 \\
3 & 1 & $0^*$ & $1^*$ & 0 & 0 & 0 & 0 & -1 & -1 & -2 & -1 \\
5 & 1 & $0^*$ & $1^*$ & 1 & 0 & 0 & -1 & -1 & -2 & -2 & -1 \\
10 & 2 & $0^*$ & $1^*$ & $2^*$ & 1& 0& -1& -2 & -3 & -2 & -1 \\
\hline
\end{tabular}

\vspace{0.5cm}
\end{center}
Table 1.- Values of $Q_i$ associated to the $N=10$
eigenstates $E_i$ for a equally spaced model
with $g=1$, $h=1,3,5,10$
and $\vep_j = 2 j - N - 1 \;(d=1)$.
The states are ordered increasingly with the
energy. The symbol $Q^*$ means that it is a bound
state, namely a state whose energy $E_i$ is smaller
that the lowest single pair energy $\vep_1= -9$.
The second column gives the number or RG cycles
as computed with eq. (\ref{rg12}).

\section{The many body problem}

The results of the previous sections strongly suggest the  existence
of a new type of excitations above the ground state of the RD model
carrying  non vanishing values of the Bethe numbers $Q_a$. In the
thermodynamic limit the BAE equation (\ref{b9}) becomes a
Richardson like eq.(\ref{b10})  which depends explicitly on the
$Q's$, namely

\beq
\frac{1}{G_0} + \frac{\pi Q_a}{\eta}
+ \sum_{j=1}^N \frac{1}{E_a - \vep_j}
- \sum_{b=1 (\neq a)}^M \frac{2}{ E_a - E_b}
= 0, \qquad a=1, \dots, M
\label{exc1}
\eeq
where $1/G_0 = \alpha/\eta$. In the rest of this section
we shall study numerically and analytically
the solutions of (\ref{exc1}) for
the equally spaced model with energy levels
$\vep_j = ( 2 j - N-1) \;\;(d=1)$.

\subsection{Numerical solutions}

The ground state of the RD model
is given by the choice $Q_a =0, \; \forall a$.
In the large N limit and in the strong coupling regime
(i.e. $g_0 = G_0/d   > g_c = 1.13459$),
the set of roots $E_a$ condense into an open arc $\Gamma$
in the complex energy plane whose end points
are given by $a= \vep_0 - i \Delta$ and
 $b= \vep_0 + i \Delta$, where $\vep_0$ is the chemical
potential and $\Delta$ is the value of the gap \cite{largeN}.
If the coupling $g_0$ is smaller than $g_c$,
there is a fraction of roots that are real
while the other ones are complex and form the arc.

Next we present several numerical solutions of eq.(\ref{exc1})
for a system with $N=100$ and $200$ energy levels
at half filling $M=N/2$.
The BCS coupling constant is fixed to $g_0=2$, so that
all the roots for the ground state form complex conjugated
pairs, except eventually for a single root which will be real.

\vspace{0.5cm} 

\begin{figure}[t!]
\begin{center}
\includegraphics[width= 10 cm,angle= 0]{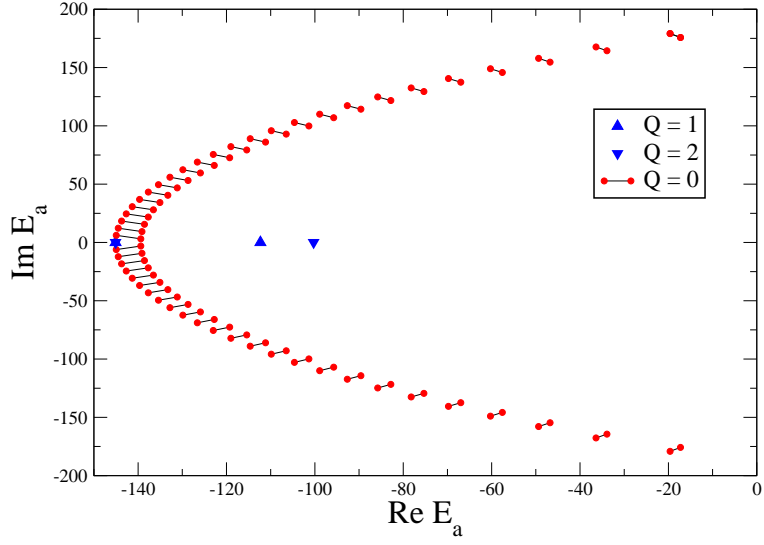}
\end{center}
 \caption{Plot of the roots $\{ E_a \}_{a=1}^M$
of eq.(\ref{exc1}) for the three cases described in the
text and the values $N = 100, M = 51, g_0 = 2,  h=2$.
The roots with $Q_a=0, \forall a$ form the ``ground state'' arc which
shrinks when $Q_1=1$ or 2 (the roots forming the two arcs
are related by the links $\bullet - \bullet$). The real
root $\xi_1$ for $Q=2$ is close to the
bottom of the energy band $\omega = -N d = -100$. 
All the energies in figs.4-7 are given in units of $d$.
}
\label{onerealQ1Q2bis}
\end{figure}

{\bf \underline{One pair:}} Figure \ref{onerealQ1Q2bis}  shows the solutions
$E_a$ corresponding to three choices: i) $\{ Q_a = 0\}_{a=1}^M$,
ii) $Q_1=1$ and   $\{ Q_a = 0\}_{a=2}^{M}$ and
iii) $Q_1=2$ and   $\{ Q_a = 0\}_{a=2}^{M}$.
For the choice i) the $M$ roots $E_a$ form the largest arc
located to the left, 
while for ii) and iii)  the $M-1$ roots with $Q_a=0$
form a smaller arc, while the real root with $Q_1=1$
or 2 lie on the real axis far apart from the arc.
We have found more than one solution with $Q_1=1$ or 2 which 
must correspond
to higher excited states of the one Copper pair problem,
as shown in the previous section.
In table 2 we collect
the  numerical and theoretical
values of the real root $E_1  \equiv \xi_1$ for several
values of the ratio $Q_1/h$, as well as the excitation energy
of the state. The theoretical values will be obtained
in the next subsection.

\vspace{1cm}

\begin{center}
\begin{tabular}{|c|c|c|c|c|c|c|}
  \hline
  $Q_1/h$ & $\xi_1^{\rm num}$ & $\xi_1^{\rm th}$ & $\xi_1^{\rm th'}$
& $E$ &  ${E}^{\rm num}_{\rm exc}$ & ${E}^{\rm th'}_{\rm exc}$\\ \hline
 3/2 & -97.365 & -100.021 & -100.021 & -2694.635 & 217.998 & 216.405 \\
{\bf 1}    &{\bf -100.277} & 
{\bf -100.477} &{\bf  -100.481} & 
{\bf -2693.203} & {\bf 219.43}  & {\bf 216.618} \\
{\bf  1/2} &{\bf -112.317} & 
{\bf -111.941} &{\bf  -112.059} & 
{\bf -2686.975} & {\bf 225.658}  & {\bf 222.225} \\
 1/3 &-132.24  & -140.931 & -141.476 & -2697.788 & 214.845 & 238.416 \\
  \hline
\end{tabular}
\end{center}

Table 2.- Numerical and theoretical values of the real
root $\xi_1$, total energy $E$
and excitation energy $E_{\rm exc}= E- E_0$ ($E_0$ 
is the ground state energy)  
for several values of $Q_1/h$. The rest of the parameters
are fixed to  $N = 100, M = 51, g_0 = 2$. The cases
depicted in fig. \ref{onerealQ1Q2bis}, 
namely $h=2, Q_{1}=1$ and 2, 
correspond to the values $Q_1/h= 0.5$ and 1 in bold face. 

\vspace{0.5cm}

\begin{figure}[t!]
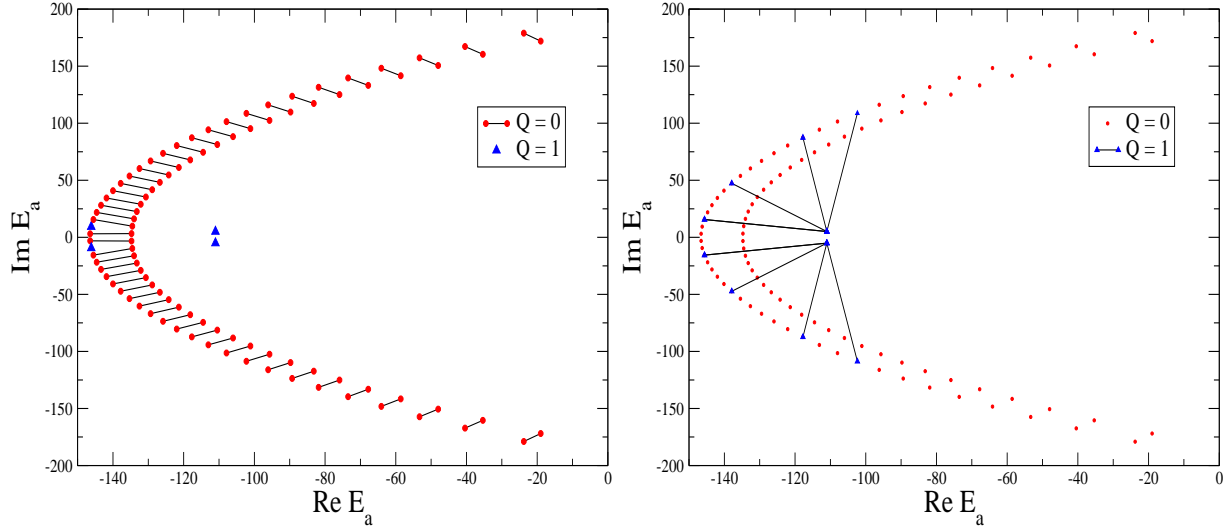

\begin{center}
\hbox{\includegraphics[height= 7 cm, width= 8 cm]{twocomplexQ1bis}
\includegraphics[height = 7 cm, width= 8 cm]{N100Q1_together}}
\end{center}
\caption{Left: Roots $E_a$
of eq.(\ref{exc1}) for
the values $N = 100, M = 50, g_0 = 2,  h=2$.
There are two roots $\xi_{1,2}$ 
with $Q_{1,2}=1$ forming a complex conjugated pair. Right:
Different choices of the roots for which $Q_a=1$ yield the same result.
}
\label{twocomplex}
\end{figure}

{\bf \underline{Two pairs:}} Fig.~\ref{twocomplex}-left shows the solutions
$E_a$ corresponding to two choices: i) $\{ Q_a = 0\}_{a=1}^M$
and ii) $Q_1= Q_2 = 1$ and  $\{ Q_a = 0\}_{a=3}^M$
In the second case the roots $E_{1,2} \equiv \xi_{1,2}$ 
form a complex conjugated pair near the real axis.
Observe how the $M$ roots of the ground state arc
reorganize themselves into a new arc with two less roots
having $Q_{1,2}=1$. In the numerical program
one can choose another pair of roots, say $E_{7}$ and  $E_{8}$,
getting the same result. This is shown in Fig~\ref{twocomplex}-right. 
In table 3 we collect
the  numerical and theoretical
values for the complex roots $E_{1,2} \equiv \xi_{1,2}$ for several
values of the ratio $Q_{1,2}/h$, as well as the excitation energy
of the state. The theoretical values will be obtained
in the next subsection.

\vspace{0.5cm}

\begin{center}
\begin{tabular}{|c|c|c|c|c|c|c|}
  \hline
$Q_{1,2}/h$&$\xi^{\rm num}_{1,2}$&$\xi^{\rm th}_{1,2}$
&$E$& $E^{\rm num}_{\rm exc}$& ${E}^{th}_{\rm exc}$ \\ \hline
  0.83 & $-100.29  \pm  1.515 \;i$ & $-100.092 \pm  1.339 \;i$ & -2481.341 & 432.458 & 432.869 \\
 {\bf 0.50} & ${\bf -111.008 \pm  5.058 \;i}$ & 
${\bf -110.632 \pm  4.963 \;i}$ & 
{\bf -2470.368} &{\bf  443.431} &{\bf  442.935}  \\
  0.38 & $-126.262 \pm 11.498 \;i$ & $-126.457 \pm  8.12 \;i$ & -2464.325 & 449.474 & 459.445 \\
  \hline
\end{tabular}
\end{center}

Table 3.- Numerical and theoretical values of the complex
roots $E_{1,2} \equiv \xi_{1,2}$, total energy $E$
and excitation energy $E_{\rm exc}$
for several ratios $Q_{1,2}/h$. The rest of the parameters
are fixed to  $N = 100, M = 50, g_0 = 2$. The case
depicted in fig. \ref{twocomplex}, namely $h=2, Q_{1,2}=1$,
corresponds to the value $Q_{1,2}/h= 0.5$ in bold face.

\vspace{1cm} 

\begin{figure}[t!]
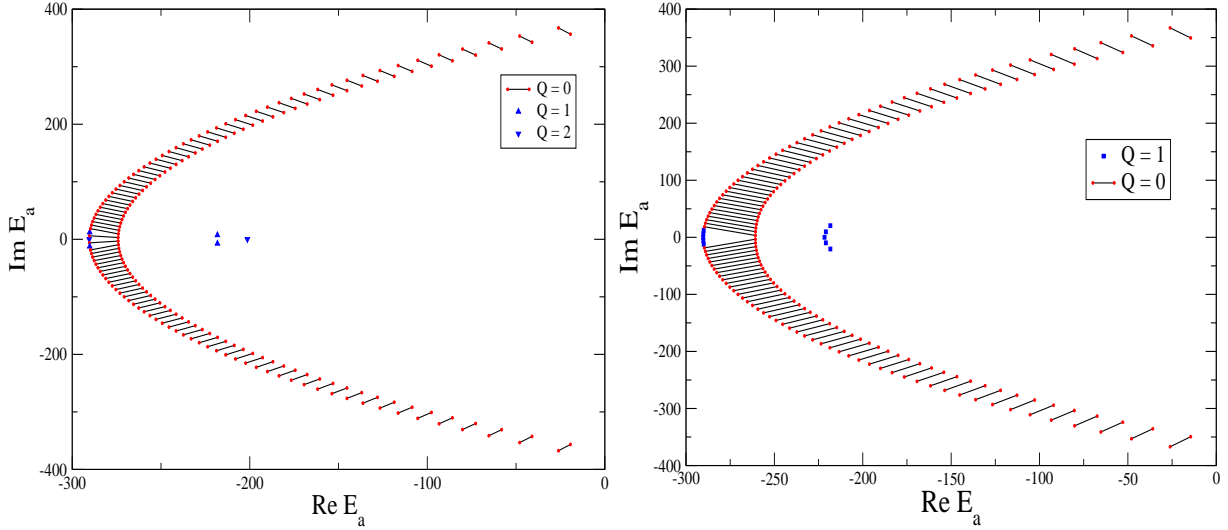

\begin{center}
\hbox{\includegraphics[height= 7 cm, width= 8 cm]{threepairs}
\includegraphics[height = 7 cm, width= 8 cm]{fivepairsQ1}}
\end{center}
\caption{
Roots $E_a$
of eq.(\ref{exc1}) for the cases: 
Left: $N = 200, M = 100, g_0 = 2,  h=2$. There are two complex roots with
$Q_{1,2}=1$ (the closest to the arcs) and a real root with
$Q_{3}=2$ (the farther from the arcs).
Right:  $N = 200, M = 101, g_0 = 2,  h = 4$.
There are five roots with $Q_a=1$.
}
\label{three-five}
\end{figure}

{\bf \underline{Three pairs:}} Fig.~\ref{three-five}-left shows a solution with
three roots formed by a complex pair with $Q_{1,2} = 1$
and a real pair with $Q_{3} =2$. In table 4 we collect the results. 
\begin{center}
\begin{tabular}{|c|c|c|c|c|c|c|}
  \hline
$\xi^{\rm num}_{1,2}$&$\xi^{\rm num}_{3}$& 
$\xi^{\rm th}_{1,2}$&$\xi^{\rm th}_{1,2}$& 
$E^{\rm num}_{\rm exc}$& ${E}^{th}_{\rm exc}$ \\ \hline
$-218.229 \pm 7.365 i$ &
-201.356 & 
$-222.578 \pm 7.115 i$ &
-200.949 &
1319.06 &
1320.5 \\
\hline
\end{tabular}
\end{center}

Table 4.- Numerical and theoretical values associated to 
fig.~\ref{three-five}-left. 

\vspace{0.5cm} 

{\bf\underline{Five pairs:}} Fig.~\ref{three-five}-right shows
a solution with five roots with
$Q_{1,\dots,5}=1$, one of which is real and the remaining four
are two complex conjugated pairs.
As can be seen they form a small arc which one  expects to become
larger with the number of $Q_a=1$ roots.

\vspace{0.5cm} 

\begin{figure}[t!]
\begin{center}
\includegraphics[width= 10 cm,angle= 0]{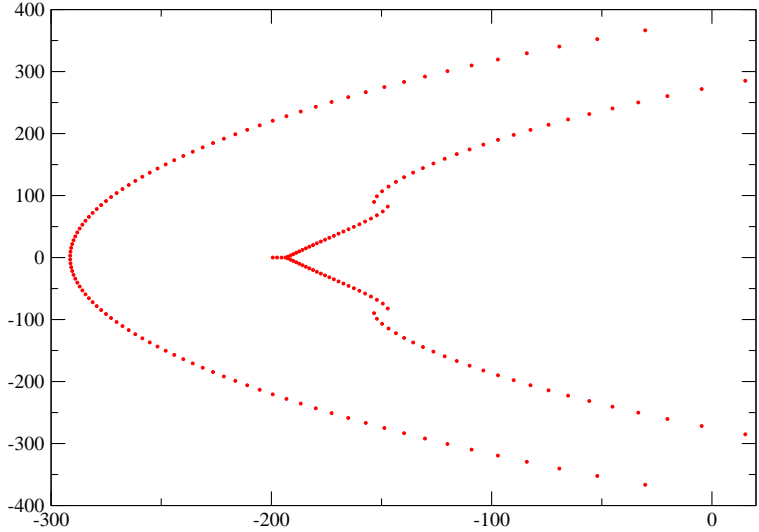}
\end{center}
\caption{
Roots $E_a$
of eq.(\ref{exc1}) for
the values $N = 200, M = 100, g_0 = 2,  h = 4$.
The wider arc on the LHS is the usual ground 
state arc.
The Y shaped arc in the middle corresponds to $M/2 = 50$
roots with $Q_a=1$, while the two disconnected branches 
above and below it contain the remaining 50 roots
with $Q_a=0$. 
}
\label{four_arcs}
\end{figure}

{\bf \underline{M/2 pairs:}} Fig.~\ref{four_arcs} shows a
solution with $M=100$ roots where half
of them have $Q_a=0$ and the other half
have $Q_a=1$. This represents
a high energy state which is intermediate
between the $Q=0$ and $Q=1$ states
where all the roots condense into single arcs.

The main conclusions we can draw from the previous numerical
results, and others not shown above,
are the following. In the thermodynamic limit
where the number of roots $E_a$
with $Q_a > 0$ is kept finite as $N$ and $M$ become
very large we have:

\begin{itemize}

\item The roots with $Q_a=0$ form an open arc $\tilde{\Gamma}$
which is a slight perturbation of the arc $\Gamma$ formed
by the $M$ roots of the ground state.

\item The roots with a non vanishing value of $Q_a$ fall
into arcs characterized by $Q_a$. 

\end{itemize}

\subsection{Analytic solution}

The previous results suggest the steps to follow in the analytical
study of eq.(\ref{exc1}). We shall employ the methods of complex
analysis developed to establish the connection between the
Richardson eqs. in the thermodynamic limit and the mean field BCS
solution \cite{G-book,largeN} (se appendix A for a review).

Let us first rewrite eq.(\ref{exc1}) as follows,

\beq
 \sum_{j=1}^N \frac{1/2}{\vep_j - E_a}
- \sum_{b=1 (\neq a)}^M \frac{1}{ E_b - E_a} - \frac{1}{2 G_a}
= 0
\label{d1}
\eeq
where
\beq
 \frac{1}{G_a} =  \frac{1}{ G_0} +  \frac{\pi Q_a}{\eta},
\qquad
 \frac{1}{G_0} =
\frac{1}{\eta} \arc \left( \frac{\eta}{G} \right)
\label{d2}
\eeq
and let us suppose that $Q_a = 0$ for all roots  $E_a$
except for a finite
number in the limit $N\rightarrow \infty$. We shall call
the latter roots
$\xi_\alpha \; (\alpha =1, \dots, D)$.  Making an electrostatic
analogy, 
eqs.(\ref{d1}) and (\ref{d2}) imply that on
each root $E_a$  acts an electric field $-1/2 G_a$, which
depends on the value of $Q_a$. The
roots with $Q_a =0$ see an
electric  field $-1/2 G_0$, while
the $Q_a > 0$ roots see a stronger field.
Assuming that all the roots $Q_a=0$ form
a single arc   $\tilde{\Gamma}$, then
the roots $\xi_\alpha$, with $Q_a > 0$,
must lie outside  $\tilde{\Gamma}$, namely

\beq
E_a =
\left\{
\begin{array}{cc}
\xi \in \tilde{\Gamma}, & Q_a = 0 \\
\xi_\alpha \notin  \tilde{\Gamma}, & Q_a > 0
\\
\end{array}
\right.
\label{d2bis}
\eeq

There may also exist roots with $Q_a=0$ lying outside
the arc $\tilde{\Gamma}$. They will be considered later on.
In the continuum limit eqs.(\ref{d1}) split into
two sets of eqs. (see appendix A for definitions),

\barray
& & \int_\Omega \frac{ d \vep \;  \rho(\vep)}{\vep - \xi}
- \sum_{\alpha = 1}^D \frac{1}{ \xi_\alpha - \xi}
- P \int_{\tilde{\Gamma}} \frac{ |d \xi'|\;
\tilde{r}(\xi')}{ \xi' - \xi}
- \frac{1}{2 G_0} =  0,
\qquad \xi\in \tilde{\Gamma},
\label{d3} \\
& & \int_\Omega \frac{ d \vep \;  \rho(\vep)}{\vep - \xi_\alpha}
- \sum_{\beta = 1 (\neq \alpha)}^D \frac{1}{ \xi_\beta - \xi_\alpha}
- P \int_{\tilde{\Gamma}} \frac{ |d \xi'|\;
\tilde{r}(\xi')}{ \xi' - \xi_\alpha}
- \frac{1}{2 G_\alpha} =  0, \qquad \xi_\alpha \notin \tilde{\Gamma}
\label{d4}
\earray

\no
where the first equation holds for the roots $\xi$ on the arc
$\tilde{\Gamma}$
and the second equation holds for the roots $\xi_\alpha$
outside  $\tilde{\Gamma}$.
The density
of roots $\tilde{r}(\xi)$ along the arc $\tilde{\Gamma}$,
and the total energy of the state
$E$ are given by (see  eqs.(\ref{c3},\ref{c4}))

\barray
& & \int_{\tilde{\Gamma}}  |d \xi| \;  \tilde{r}(\xi)  =  M-D,
\label{d5} \\
& &
\int_{\tilde{\Gamma}}   |d \xi| \;   \xi \;  \tilde{r}(\xi)
+ \sum_{\alpha=1}^D \xi_\alpha
=  E. \label{d6}
\earray

\no
The solution of eqs.(\ref{d3},\ref{d4}) follows the same steps
as in Appendix A. Let us summarize
the results. The density function $\tilde{r}(\xi)$ can
be found from a formula similar to (\ref{c6})
with a modified electric field $\tilde{h}(\xi)$ given
by,

\beq
\tilde{h}(\xi) = \tilde{R}(\xi) \;
\left( \int_\Omega
\frac{ \rho(\vep)}{\tilde{R}(\vep)} \; \frac{ d \vep }{ \vep - \xi}
- \sum_{\alpha=1}^D \frac{1}{\tilde{R}(\xi_\alpha) (\xi_\alpha - \xi)}
\right),
\label{d7}
\eeq

\no
with

\beq
\tilde{R}(\xi) = \sqrt{ (\xi - \tilde{a})(\xi -\tilde{b})}
\label{d8}
\eeq

\no
and where $\tilde{a}= \tilde{\vep}_0 - i \tilde{\Delta},
\tilde{b} = \tilde{\vep}_0 + i \tilde{\Delta}$ are
the end points of the arc $\tilde{\Gamma}$.
Using eq.(\ref{d7}) into (\ref{d3}) one gets after
some algebra,

\beq
 \frac{1}{2 G_0} =
\int_\Omega  \frac{d \vep \; \rho(\vep)}{
 \sqrt{(\vep - \tilde{\vep}_0)^2 + \tilde{\Delta}^2}}
-  \sum_{\alpha=1}^D \frac{1}{
 \sqrt{(\xi_\alpha - \tilde{\vep}_0)^2 + \tilde{\Delta}^2}}
\label{d9}
\eeq

\no which is a modified
gap equation analogous  to (\ref{c13}).
 Similarly (\ref{d5}) gives the modified
chemical potential eq. (see (\ref{c17})),

\begin{equation}
M - D =  \int_\Omega  d \vep \; \rho(\vep) \;  \left(
1 - \frac{ \vep - \tilde{\vep}_0}{
 \sqrt{(\vep - \tilde{\vep}_0)^2 + \tilde{\Delta}^2}}
\right) - \sum_{\alpha=1}^D \left( 1 -
 \frac{\xi_\alpha - \tilde{\vep_0}}{
 \sqrt{(\xi_\alpha - \tilde{\vep}_0)^2 + \tilde{\Delta}^2}}
\right)
\label{d10}
\end{equation}

\no
These two equations determine the end points
$\tilde{\vep}_0 \pm  i \tilde{\Delta}$
of the arc $\tilde{\Gamma}$. They are formally equivalent
to the eqs.(\ref{c13}) and (\ref{c17}) if one defines
an effective density

\beq
\tilde{\rho}(\vep) = \rho(\vep) + \delta \rho(\vep)
, \qquad \delta \rho(\vep) = - \sum_{\alpha =1}^D
\delta(\vep - \xi_\alpha)
\label{d13}
\eeq

Similarly the eq.
(\ref{d4}) for the roots $\xi_\alpha$ turns into,

\beq
 \frac{ \pi Q_\alpha}{2 \eta} =
 \int_\Omega  d \vep \;
\frac{\rho(\vep)}{ \vep - \xi_\alpha}
\frac{\tilde{R}(\xi_\alpha)}{\tilde{R}(\vep)}
- \sum_{\beta = 1 (\neq \alpha)}^D
\frac{1}{\xi_\beta - \xi_\alpha}
\frac{\tilde{R}(\xi_\alpha)}{\tilde{R}(\xi_\beta)}
+ \frac{\tilde{R'}(\xi_\alpha)}{\tilde{R}(\xi_\alpha)}
\label{d11}
\eeq

\no where the RHS can be identified with the electric field
$\tilde{h}(\xi_\alpha)$ (see eq.(\ref{d7})), after subtracting the
pole at $\xi = \xi_\alpha$. Finally, the energy of the state can be
derived from eq.(\ref{d6}) and it reads,

\barray
 E &  = &
- \frac{\tilde{\Delta}^2}{ 4 G_0} +
\int_\Omega d \vep \;  \vep \;   \left(
1 - \frac{ \vep - \tilde{\vep}_0}{
 \sqrt{(\vep - \tilde{\vep}_0)^2 + \tilde{\Delta}^2}}
\right) \rho(\vep)
\label{d12} \\
& &
+ \sum_{\alpha=1}^D
\xi_\alpha
-  \sum_{\alpha=1}^D \xi_\alpha
\left( 1 - \frac{ \xi_\alpha - {\tilde{\vep}_0}}{
 \sqrt{(\xi_\alpha - \tilde{\vep}_0)^2 + \tilde{\Delta}^2}}
\right)
\nonumber
\earray

\no

Doing a computation similar to the one that leads to 
eqs.(\ref{c23}) and (\ref{c25})
in Appendix B
one gets the excitation energy of the state
$E_{exc} = E- E_0$ ($E_0$ is the ground state energy)

\beq
E_{exc} = \sum_{\alpha =1}^{D}
\sqrt{ (\xi_\alpha - \vep_0)^2 + \Delta^2} \;,
\label{d13bis}
\eeq
where $\Delta$ and $\vep_0$  are the unperturbed
values of the gap and chemical potentials given by
the solution of eqs.(\ref{c13}) and (\ref{c17}) with
$G=G_0$.

\subsection{Analytic versus numerics}

Let us compare the analytic and the numerical results
obtained previously. For one real pair $\xi_1$
the eq.(\ref{d11}) becomes,

\beq
 \frac{ \pi Q_1}{2 \eta} =
 \int_\Omega  d \vep \;
\frac{\rho(\vep)}{ \vep - \xi_1}
\frac{\tilde{R}(\xi_1)}{\tilde{R}(\vep)}
+ \frac{\tilde{R'}(\xi_1)}{\tilde{R}(\xi_1)}
\label{d14}
\eeq
In the large N limit we can approximate $\tilde{R}(\xi)$
by $R(\xi)$, so that (\ref{d14}) simplifies,
\beq
 \frac{ \pi Q_1}{2 \eta} = h(\xi_1) +
\frac{\xi_1 - \vep_0}{( \xi_1 - \vep_0)^2 + \Delta^2}
\label{d15}
\eeq
where we have used (\ref{c15}). In the equally spaced
model at half filling the integral giving $h(\xi)$
is given by the formula \cite{largeN},
\begin{equation}
h(\xi) = \frac{1}{4 d}  \;
\log{ \frac{
(\xi - \omega)( \Delta^2 - \xi \omega +
\sqrt{ (\Delta^2 + \xi^2)(\Delta^2 + \omega^2)}) }{
(\xi + \omega)( \Delta^2 + \xi \omega +
\sqrt{ (\Delta^2 + \xi^2)(\Delta^2 + \omega^2)})}} \;.
\label{d16}
\end{equation}
To compare with the numerical results collected in table 2
we solve eq.(\ref{d15}) for the values,
$N = 100, \; g_0 = 2, \;  d=1$ which implies that
$\Delta = N d/\sinh(1/g_0) = 191.903$, $\vep_0 = 0$
and $\omega = d N = 100$. The result is given in table 2 in
the column $\xi_1^{\rm th'}$.
The excitation energy is obtained
using (\ref{d13bis}), namely
$E_{\rm exc} = \sqrt{ (\xi_1^{\rm th'})^2 + \Delta^2}$ and the 
values appear in the column $E^{\rm th'}_{\rm exc}$ of table 2.
Observe the good agreement between  the numerical
and theoretical values.

As can be seen from eq.(\ref{d16}) the field $h(\xi_1)$
is proportional to $1/d \sim N$, hence in the large N limit
the eq. (\ref{d15}) can be approximated by

\beq
 \frac{ \pi Q_1}{2 \eta} = h(\xi_1)
\label{d17}
\eeq
On the other hand the change of variables \cite{largeN}
\beq
\xi = \frac{i \Delta \cosh u}{
\sqrt{1 - \left( \frac{ \Delta}{\omega} \sinh u \right)^2}}
\label{d18}
\eeq
leads to a very simple expression for $h(\xi)$, namely
\beq
h(\xi) = - \frac{u}{2 d}
\label{d19}
\eeq
So that the solution of eq.(\ref{d17}) is given simply by
\beq
\xi_1 = - \frac{\Delta \cosh( \pi Q/h) }{
\sqrt{- 1 + \left( \frac{ \Delta}{\omega}
\sinh(\pi Q/h) \right)^2}}
\label{d20}
\eeq
The column $\xi_1^{\rm th}$ of table 2 collects
several values of (\ref{d20}), which are quite close
to the values of $\xi_1^{\rm th'}$. The solution
(\ref{d20}) remains real and below the bottom
of the band, i.e. $\xi_1 <-  \omega$, 
whenever $\pi Q/h > 1/g_0$. 

To check the results of table 3 one has to solve eq.(\ref{d11})
for two complex roots, say $\xi_1$ and $\xi_2 = \xi_1^*$ with
$Q_1 = Q_2$,
\beq
 \frac{ \pi Q_1}{2 \eta} = h(\xi_1) -
\frac{1}{\xi_2 - \xi_1} \frac{R(\xi_1)}{R(\xi_2)}, \qquad
   \frac{ \pi Q_2}{2 \eta} = h(\xi_2) -
\frac{1}{\xi_1 - \xi_2}  \frac{R(\xi_2)}{R(\xi_1)}
\label{d21}
\eeq
Notice that we drop the term $R'(\xi)/R(\xi)$ in these equations.
The numerical results are shown in the column $\xi^{\rm th}_{1,2}$.
Similarly the excitation energy is given by
$E_{\rm exc} = 2 Re[\sqrt{ (\xi_1^{\rm th})^2 + \Delta^2}]$
Finally, the theoretical results of table 4 are obtained
by choosing $\xi_{1,2}^{\rm th}$ as the solutions of eqs.(\ref{d21}),
and $\xi_{3}^{\rm th}$ from eq.(\ref{d20}). Strictely
speaking one must solve eqs.(\ref{d11}) for three
roots with  $Q_a =1,1,2$, but they can be approximated
as above. The energy
is given by (\ref{d13bis}) summing over the three roots
$\xi_i^{\rm th}$.

\subsection{The $Q$-excitations in the PBCS ansatz}

Let us summarize the results obtained so far. Using the grand canonical
BCS ansatz  we showed the existence of different solutions of the
gap equation in the RD model characterized by an integer $Q \geq 0$,
where the $Q=0$ solution corresponds to the ground state and the
solutions with $Q \geq 1$ correspond to higher excited states with
different condensation energies. Later on, the exact solution of the
model led us to identify  the integer $Q$ with the Bethe numbers
appearing in the solution of the BAE for the $M$ roots, namely $Q=
Q_a, \;\forall a$. This suggested the existence of low lying excited
states for which $Q_a=0$ for most of the roots, in the thermodynamic
limit, except for a finite number of them, for which $Q_a \geq 1$.
These new type of excitations cannot be derived from the standard
mean field analysis in the grand canonical 
 ensemble which are given in terms of
the familiar Bogoliubov quasiparticles. This is why we have to use
the exact solution to find them. One may wonder however if there is
an alternative way to derive the $Q-$excitations, which could be
valid in more general circumstances. 
We now  argue that this is indeed
possible using the projected BCS ansatz. Let us write the BCS state
defined in eq.(\ref{a3}) as
\beq
|\psi_0 \rangle \propto {\rm exp}\left(
\sum_k g_k b^\dagger_k \right)  \; |0 \rangle
\label{q1}
\eeq
where $g_k = v_k/u_k$ is the wave function
of the Cooper pair in momentum space.
A state with $M$ pairs can be obtained
from (\ref{q1}) keeping the
$M^{\rm th}$-term of the Taylor expansion, i.e.
\beq
|M \rangle \propto \left(
\sum_k g_k b^\dagger_k \right)^M  \; |0 \rangle
\label{q2}
\eeq
which is called the Projected BCS state (PBCS). While the grand canonical
 BCS
state (\ref{q1}) is an eigenstate of the phase operator we see that
(\ref{q2}) is an eigenstate of the electron number operator. In the
large $N$ limit the results obtained with both ansatze agree to
leading order in $N$. However for finite values of $N$ the BCS and
PBCS give different results as ocurred when they were 
applied to the study of
ultrasmall superconducting grains \cite{BvD,DMRG1,DMRG2}. In the RD
model the PBCS state corresponding to
 the $Q^{\rm th}$-solution can be written as
\beq
|M_Q \rangle \propto \left(
\sum_k g_k^{(Q)} b^\dagger_k \right)^{M_Q}  \; |0 \rangle
\label{q3}
\eeq
where $g^{(Q)}_k$ is the associated wave function.
The ground state is given by
(\ref{q3}) with $Q=0$.
We shall conjecture that the states found in the subsections
V-A and V-B can be approximated by the following PBCS states
\beq
|M_0, M_1, \dots \rangle \propto \left(
\sum_k g_k^{(0)} b^\dagger_k \right)^{M_0}
 \left(
\sum_k g_k^{(1)} b^\dagger_k \right)^{M_1}
\dots
  \; |0 \rangle
\label{q4}
\eeq
where  $M_0 \sim {\cal O}(M)$ 
is the number of roots with $Q_a=0$, $M_1 \sim {\cal O}(1)$
the number of roots with $Q_a=1$, etc. Eq.(\ref{q4}) certainly holds
in the case where the operators $b_k$ are ordinary bosons instead
of hard core ones. In this case the RD Hamiltonian can be
diagonalized by a canonical transformation of the boson operators
$b_k$ which amounts to solving  the BAE for one pair,
eq.(\ref{rg1}). Each term in (\ref{q4}) would correspond to the
bound state solutions of the one pair problem with $Q=0, 1, \dots$.

An important consequence of this discussion is that the Q-excitations
must behave as ordinary bosons. In the canonical picture one simply
adds one more pair to the $Q > 0$ states, which then increases the 
excitation energy. In the exact solution this corresponds to 
removing pairs from the ground state arc and place them into
the small arcs with $Q_a > 0$. 
This property is in sharp contrast with
the standard quasiparticles which are fermions.

\subsection{Pair-hole excitations as $Q_a=0$ states}

\begin{figure}[t]
\begin{center}
\includegraphics[width= 10 cm,angle= 0]{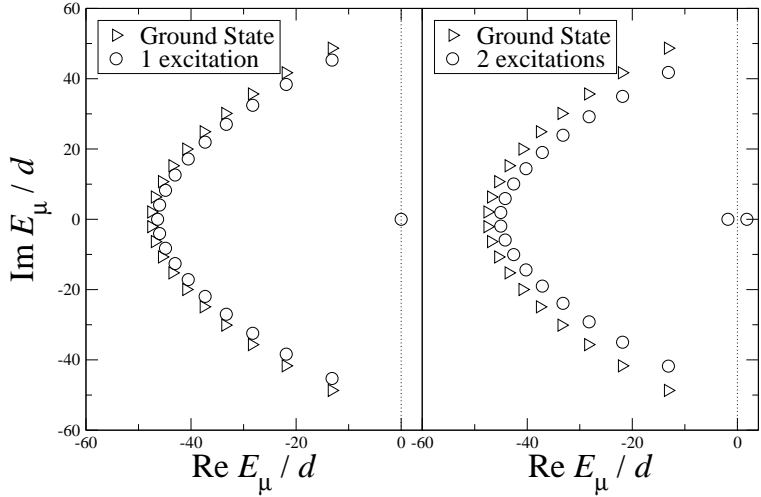}
\end{center}
\caption{Position $M=20$ pairs of the states at $g=1.5$.
The arcs $\tilde{\Gamma}$ for
19 pairs (left) and 18 pairs (right)
are a slight modification of the ground state
arc $\Gamma$
(20 pairs)  \cite{excita}.
}
\label{excitfig2}
\end{figure}

So far we have considered
the solutions of eq.(\ref{exc1})
where some of the $Q_a$ are non zero
which has the effect that the associated roots
come out from the ground state arc.
In \cite{excita} it was shown
that the Richardson eqs. (\ref{b7})
already contains many solutions where this happens.
The latter solutions correspond to the pair-hole
excitations of the BCS model, to be distinguished
from the excitations where the energy levels are
blocked by single electrons.
(see appendix B).
In the large $N$ limit the pair-hole excitations
are characterized by the fact that $N_G$ roots
$E_a$ lie outside the arc $\tilde{\Gamma}$
formed by the $M-N_G$ remaining ones \cite{excita}.
In figure  \ref{excitfig2} we show an example
of two excited states with $N_G=1$ and 2 roots
for a model with $M=20$ roots.
Figure  \ref{excitfig2}a  is similar
to fig.\ref{onerealQ1Q2bis} since in both cases
one root comes out of the ground state
arc $\Gamma$. They differ however in the value of
$Q_1$, which is 1 in fig. \ref{onerealQ1Q2bis}
and 0 in fig. \ref{excitfig2}a. In \cite{excita}
it was found that the roots
$E_\alpha \; (\alpha=1, \dots, N_G)$,
lying outside the arc $\tilde{\Gamma}$
satisfy in the
large $N$ limit the following eq.
\beq
0= \int_\Omega  d \vep \;
\frac{\rho(\vep)}{ \vep - \xi_\alpha}
\frac{\tilde{R}(\xi_\alpha)}{\tilde{R}(\vep)}
- \sum_{\beta \neq \alpha}^{N_G}
\frac{1}{( \xi_\beta - \xi_\alpha)}
\frac{R(\xi_\alpha)}{ R(\xi_\beta)}
+ \frac{R'(\xi_\alpha)}{ R(\xi_\alpha)}
\;,
\label{c26}
\eeq

\no
which coincides with eq.(\ref{d11}) upon the choice
$Q_\alpha =0$.
Moreover the
energy of these excited states is given by,
\beq
E_{exc} = \sum_{\alpha =1}^{N_G}
\sqrt{ (\xi_\alpha - \vep_0)^2 + \Delta^2} \;,
\label{c27}
\eeq

\no
This formula coincides with eq.(\ref{d13bis})
giving the energy of the $Q_a$-excitations.
Hence the pair-hole excitations
of \cite{excita} can be seen as $Q_a=0$ excitations.

Summarizing, in the RD model
there are three types of excitations
in the canonical ensemble, i) the ones obtained
by the blocking of energy levels, ii) the pair-hole
excitations and iii) the $Q-$excitations.
The first two have already been considered for the
standard BCS model and in the g.c. ensemble they 
correspond to the quasiparticles which have fermionic
statistics. The last ones are 
specific of the RD model and have bosonic statistics. 
These results must have profound consequences 
in the thermodynamical properties of the system
and its response to external fields.

\section{Conclusions and prospects}

In this paper we have analyzed in more depth
the Russian doll BCS model using its exact solution.
We have shown that the mean field solution
obtained in  \cite{RD} agrees in the
limit of large number of particles $N$ with the
exact solution to leading order in $N$. In doing so
we have identified the integer $Q$, that labels the mean
field solutions, with the Bethe numbers appearing in the BAE.
This integer  also has  a RG meaning since it counts the
number of RG cycles. This idea
is clarified by deriving the BAE for one Cooper pair
using the RG method of Glazek and Wilson \cite{GW}.
The numerical
solution of the latter equation
shows the appearance of bound states with $Q > 0$
and of unbounded states with $Q < 0$. We have also studied
the many body case in the large $N$ limit, where the
BAE can be approximated by Richardson like equations
where the value of the effective 
BCS coupling constant depends on the Bethe numbers $Q_a$.
We have solved numerically and analytically
these equations showing the existence of solutions where the Bethe
numbers $Q_a$ may vary with the roots. The most
interesting case is when $Q_a$ vanishes
for most of the roots except for a small number where they
are positive. These solutions correspond to low energy excitations
where Cooper pairs, forming the ground state,
jump into excited states characterized by
positive values of $Q_a > 0$.
In this sense $Q_a$  becomes a
sort of principal quantum number of the Cooper pairs.
These new type of elementary excitations should be added
to the standard ones to describe the low energy
spectrum of the model, whose thermodynamical properties
and response to external fields will be modified.

\vspace{1cm}

\noindent {\em Acknowledgments}. We would like to thank J.M.
Rom\'an, J. Dukelsky, G. Morandi and M. Roncaglia for discussions. 
This work is
supported by the grants FIRB 2002-2005 project number RBAU018472 of
Italian MIUR (AA), the  NSF of the USA (AL) and the CICYT of Spain
under the contract BFM2003-05316-C02-01 (GS). We also thank the
EC Commission for financial support via the FP5 Grant
HPRN-CT-2002-00325. \vspace{-10 pt}

\section*{Appendix A: Gaudin's electrostatic solution of Richardson equations}

\begin{figure}[t!]
\begin{center}
\includegraphics[height= 10 cm,angle= -90]{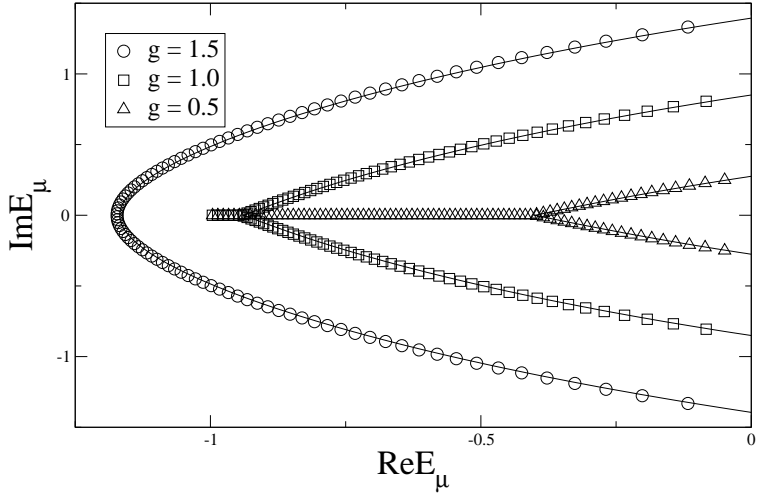}
\end{center}
\caption{Plot of the roots $E_\mu$ for the equally spaced model
in the complex plane (taken from  \cite{largeN}).
The discrete symbols denote the numerical values for $M=100$.
The continuous lines are the analytical curves
obtained in \cite{largeN}. All the energies are in units
of $\omega$.
}
\label{arcuniform}
\end{figure}

In this section we review the
electrostatic method
employed by Gaudin in the proof of the asymptotic agreement
of the exact and mean field solutions \cite{G-book}
(see also \cite{largeN,amico-electro}).
The starting point of Gaudin's approach is the observation that the
Richardson eq.(\ref{b7}), that we shall rewrite as
\beq
 \sum_{j=1}^N \frac{1/2}{\vep_j - E_a}
- \sum_{b=1 (\neq a)}^M \frac{1}{ E_b - E_a} - \frac{1}{2 G}
= 0
\label{c1}
\eeq
\no
can be seen as the equilibrium conditions for a set of
mobile charges $+1$ located at the positions
$E_a$ in the complex plane
under the effect of a
 constant electric field $-1/2G$ and
another set of charges $-1/2$ at the positions $\vep_j$.
In the large $N$ limit
the pair energy levels $\vep_j$
will be equivalent to a negative charge density
$- \rho(\vep)$ located on an interval
$\Omega = (-\omega, \omega)$
of the real axis. The total charge of this interval
is given by
\beq
- \int_\Omega d \vep \;  \rho(\vep)  = - \frac{N}{2}\;
\label{c2},
\qquad \rho(\vep) = \frac{1}{2}
\sum_{j=1}^N \delta(\vep - \vep_j)
\eeq
\no
The numerical solutions of eqs.~(\ref{c1})
are either real or complex roots in which case they
come in conjugated pairs
\cite{R-roots,largeN}.
In the large $N$ limit
the complex roots form an open arc $\Gamma$ whose end points
are given by $a = \vep_0 - i \Delta$ and
$b= \vep_0 + i \Delta$, where $\vep_0$ is
the chemical potential and $\Delta$ is
the gap (see fig.~\ref{arcuniform}). 
Let us assume for simplicity that all the
roots are complex. We shall call
$r(\xi)$
the linear charge density of roots $E_a$
along  the arc $\Gamma$.
Hence, the total number
of pairs, $M$, and energy, $E$,  are given by
\barray
\int_\Gamma  |d \xi| \;  r(\xi)  & = & M,  \label{c3} \\
\int_\Gamma   |d \xi| \;   \xi \;  r(\xi) & = & E. \label{c4}
\earray
The continuum limit of eqs.~(\ref{c1}) is
\beq
\int_\Omega \frac{ d \vep \;  \rho(\vep)}{\vep - \xi}
- P \int_{\Gamma} \frac{ |d \xi'|\;   r(\xi')}{ \xi' - \xi}
- \frac{1}{2 G} = 0,
\qquad \xi\in \Gamma,
\label{c5}
\eeq
\no
which implies the vanishing of  the total electric field
on every point of the arc $\Gamma$.
The solution of eqs.~(\ref{c5}) can be found as follows.
First of all, let us orient the arc $\Gamma$ from the point
$a$ to the point $b$, and call $L$ an anticlockwise
path encircling $\Gamma$. We look for an analytic field
$h(\xi)$ outside $\Gamma$ and the set $\Omega$, such that
\beq
r(\xi) \; |d \xi| = \frac{1}{2 \pi i}
(h_+(\xi) - h_-(\xi) ) \; d \xi,
\qquad \xi \in \Gamma,
\label{c6}
\eeq
\no
where $ h_+(\xi)$ and $ h_-(\xi)$ denote the limit
values of $h(\xi)$ to the right and left of $\Gamma$.
This can be understood using the electrostatic equivalence,
considering that the electric field presents a discontinuity proportional
to the linear charge density when crossing the arc $\Gamma$.
Next, we define a function $R(\xi)$,
with cuts along the curve $\Gamma$, by the equation
\beq
R(\xi) = \sqrt{
(\xi - a) (\xi - b)},
\label{c7}
\eeq
\no
and look for a solution which vanishes at the
boundary points of $\Gamma$, in the form
\beq
h(\xi) = R(\xi) \; \int_\Omega
 \frac{ d \vep \;  \varphi(\vep) }{ \vep - \xi}\; .
\label{c8}
\eeq
\no This field has to be constant at infinity, as can be verified
explicitly. The contour integral surrounding the charge density
$r(\xi)$ in (\ref{c5}) can be expressed as
\beq
\int_\Gamma \frac{ |d \xi'| \;  r(\xi')}{
\xi - \xi'} = \frac{1}{2 \pi i}  \int_\Gamma
 d\xi' \;
\frac{ h_+(\xi') - h_-(\xi')}{
\xi - \xi'}  = \frac{1}{2 \pi i} \int_L  d\xi' \;
\frac{ h(\xi')}{
\xi - \xi'},
\qquad \xi \in {\bf C} \Gamma,
\label{c9}
\eeq
\no where $ {\bf C} \Gamma$ is the region
outside the curve $\Gamma$. Using
eqs.~(\ref{c7}) and (\ref{c8}), one finds
for the principal value  of (\ref{c9})
\beq
P  \int_\Gamma \frac{ |d \xi'| \;   r(\xi')}{
\xi - \xi'} =
\frac{1}{2 \pi i} \int_L
\frac{ d\xi' R(\xi')}{ \xi - \xi'}
\int_\Omega \frac{ d \vep \;  \varphi(\vep)}{
\vep - \xi'}
= - \int_\Omega d \vep \;  \frac{ \varphi(\vep) R(\vep)}{
\vep - \xi}  +
\int_\Omega  d \vep \;  \varphi(\vep),
\qquad \xi \in \Gamma,
\label{c10}
\eeq
\no where we have deformed the contour of integration
$L$ into two contours, one encircling the
interval $\Omega$ (first term on the RHS) and another
one around the infinity (second term).
We are assuming that $\Gamma$, and consequently
$L$,  do not cut the interval $\Omega$, which happens
int the equally spaced model when
$g > g_c=1.13459$ \cite{largeN}.
Plugging eq.~(\ref{c10}) into (\ref{c5}),
we see that a solution is obtained provided
\barray
& &\varphi(\vep) = \frac{\rho(\vep)}{R(\vep)},
\label{c11} \\
& & \nonumber \\
& &\int_\Omega d \vep \;  \frac{\rho(\vep)}{R(\vep)}
= \frac{1}{2 G}\; ,
\label{c12}
\earray
\no
Using (\ref{c7}) the  eq.(\ref{c12}) becomes,
\beq
\int_\Omega  \frac{d \vep \; \rho(\vep)}{
 \sqrt{(\vep - \vep_0)^2 + \Delta^2}}
 = \frac{1}{2 G}\; .
\label{c13}
\eeq
\no which is nothing but the BCS gap eq.(\ref{a9}) in the
appropriate normalizations (see eq.(\ref{c19}) below). The field
$h(\xi)$ gives the charge density $r(\xi)$,
\beq
r(\xi) = \frac{1}{\pi} |h(\xi)|\; ,
\qquad \xi \in \Gamma,
\label{c14}
\eeq
\no
and its value is given by replacing
(\ref{c11}) into (\ref{c8}), i.e.\
\beq
h(\xi) = R(\xi) \; \int_\Omega
\frac{ \rho(\vep)}{R(\vep)} \; \frac{ d \vep }{ \vep - \xi}\; ,
\label{c15}
\eeq
\no
The equation fixing the arc $\Gamma$
are the equipotential curves of the total
distribution,
\beq
{\cal R} \int_{a}^\xi  d \xi' \;   h(\xi') = 0,
\qquad \xi \in \Gamma.
\label{c16}
\eeq
\no Similarly, eq.~(\ref{c3}) becomes
the chemical potential equation
\begin{equation}
M = \frac{1}{2 \pi i} \int_L d \xi \;  h(\xi)
= \int_\Omega  d \vep \; \rho(\vep) \;  \left(
1 - \frac{ \vep - \vep_0}{
 \sqrt{(\vep - \vep_0)^2 + \Delta^2}}
\right),
\label{c17}
\end{equation}
\no
while eq.~(\ref{c4}) gives the BCS expression
for the ground state energy,
\beq
 E  = \frac{1}{2 \pi i} \int_L d \xi \;  \xi\;  h(\xi)
= - \frac{\Delta^2}{ 4 G} +
\int_\Omega d \vep \;  \vep \;   \left(
1 - \frac{ \vep - \vep_0}{
 \sqrt{(\vep - \vep_0)^2 + \Delta^2}}
\right) \rho(\vep).
\label{c18}
\eeq
\no
Comparing these equations with the
corresponding ones in the BCS theory,
we deduce the following relations
between  $\Delta$, $\vep_0$, and $\rho(\vep)$,
and $\Delta_{BCS}$ (BCS gap),
$\mu$ (chemical potential), and
$n(\epsilon)$ (single particle energy density):
\beq
\Delta = 2 \Delta_{BCS}, \qquad
\vep_0 = 2 \mu, \qquad
\rho(\vep) = \frac{1}{4}\; n\left( \frac{\vep}{2} \right).
\label{c19}
\eeq
\no

\section*{Appendix B: Elementary excitations of BCS in the canonical ensemble}

In the  grand canonical ensemble
the excited states can be obtained
acting on the GS ansatz $|\psi_0 \rangle$  with the Bogoliubov
operators $\gamma_{j,\sigma} \; (\sigma = \pm)$
\beq
\gamma_{j_1,\sigma_1} \dots
\gamma_{j_n,\sigma_n} |\psi_0 \rangle ,
\qquad \gamma_{j, \pm} = u_j \, c_{j \pm}
\mp v_j \, c^\dagger_{j \mp},
\label{c20}
\eeq

\no
where the variational parameters $u_j, v_j$ are given by eqs.
similar to (\ref{a4}).
The excitation energy is given by (recall the factor of 2 in our
conventions as compare to the standard ones).
\beq
E_{exc} = E - E_0 = \frac{1}{2} \sum_{j} \sqrt{
(\vep_j - \vep_0)^2 + \Delta^2}
\label{c21}
\eeq
\no
We have to distinguish between two sorts of excitations:
E1)  quasiparticles occupying
different energy levels, which means that  all the $j's$
in eq. (\ref{c20}) are different,
and E2) quasiparticles occupying the same energy level, i.e.
$\gamma^\dagger_{j,+} \gamma^\dagger_{j,-}$. In the latter
case the factor
$(u_j + v_j \, b^\dagger_j)$ in the ground state is replaced
by the factor $(v_j - u_j \, b^\dagger_j)$.
The states  E2 are called ``real pairs'' to be distinguished from
the ``virtual pairs'' that build up the ground state \cite{BCS-book}

In the canonical ensemble the Bogoliubov operators do not even make
sense since they change the particle number which is fixed by
definition. It is however possible to see the correspondence of
these excitations in the canonical ensemble. The excitations of type
E1 correspond to blocking of energy levels and we discuss them next
in detail. The excitation of type E2 correspond to pair-hole
excitations and they are discussed in subsection V-E.

{\bf  Blocking of energy levels}

\begin{figure}[h!]
\begin{center}
\includegraphics[height= 4 cm]{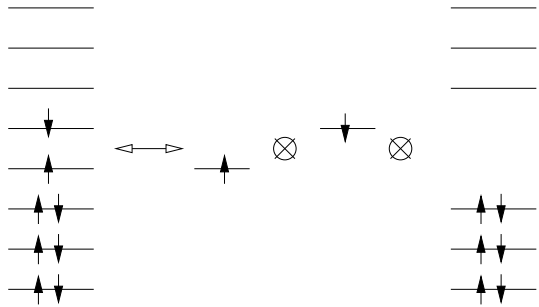}
\end{center}
\caption{Excited state obtained by breaking
a ``Cooper pair'' at $G=0$. On the RHS the
singly occupied levels are blocked and decouple
from the rest of the system.
}
\label{blocking}
\end{figure}

Let us break $P$ pairs of the
$M$ pairs forming the ground state
and place the corresponding $2P$ electrons in
different $2P$ levels belonging to the set
$B = \{j_1 < j_2< \dots < j_{2P} \}$.
All these levels are  singly occupied and therefore
the corresponding electrons  decouple from the rest
of the system contributing only with their
free kinetic energy $\frac{1}{2} \,  \vep_j, \; j \in B$.
The remaining $M-P$ pairs are then allowed to occupy
the $N- 2P$ energy levels that are left
 (see  fig.~\ref{blocking} for an example).
The problem  we are left with
is entirely similar to the original one after removing
the blocked levels. This is turn is equivalent to
consider the modified density of levels,
\beq
\tilde{\rho}(\vep) = \rho(\vep) + \delta \rho(\vep)
, \qquad \delta \rho(\vep) = -\frac{1}{2} \sum_{j \in B}
\delta(\vep - \vep_j)
\label{c22}
\eeq
where  $\rho(\vep)$ was defined in (\ref{c2}).
This small perturbation of $\rho(\vep)$ translates
into a small perturbation of
the density, i.e.
$\tilde{r}(\xi) = r(\xi) + \delta r(\xi)$
and that of the
arc $\Gamma$ whose ends points will move slightly, i.e.
$\tilde{\vep}_0 = \vep_0 + \delta \vep_0$
and $\tilde{\Delta} = \Delta + \delta \Delta$.
The eqs. for the deviations in the chemical
potential and the gap can be obtained
by taking the variation of the gap eq.(\ref{c13})
and the chemical potential eq.(\ref{c17}):
\barray
\int_\Omega d \vep \; \frac{\rho(\vep)}{R(\vep)^3} \;  \left[
\Delta \delta \Delta - (\vep - \vep_0) \delta \vep_0 \right]
& = &
\int_\Omega d \vep \;  \frac{\delta \rho(\vep)}{R(\vep)}
\label{c23}
\\
\int_\Omega d \vep \;  \frac{\rho(\vep) \Delta}{R(\vep)^3} \;  \left[
 (\vep - \vep_0)  \delta \Delta + \Delta \delta \vep_0 \right]
& = & \delta M -
\int_\Omega d \vep \;  \delta \rho \left( 1 -
\frac{ \vep - \vep_0}{R(\vep)}
\right)
\nonumber
\earray
where $R(\vep)$ is given by eq.(\ref{c7}).
Taking the variation in (\ref{c18}) one finds
\barray
\delta E & = &
- \frac{ \Delta \delta \Delta }{2 G}
+ \int_\Omega d \vep \; \vep \; \frac{  \rho (\vep) \Delta}{
R(\vep)^3} \left[ (\vep - \vep_0) \delta \Delta
+ \Delta \delta \vep_0 \right] \label{c24}
\\
& & + \int_\Omega d \vep \; \vep  \;
\delta \rho(\vep) \left( 1 - \frac{ \vep - \vep_0}{
R(\vep)} \right)
\nonumber
\earray
Using the gap eq. for the first term in
the RHS and applying the eqs. (\ref{c23})
one gets,
\barray
\delta E & = & \vep_0 \delta M
+ \int_\Omega d \vep \; \delta \rho(\vep)
 \; \left[ \vep - \vep_0 - R(\vep) \right]
\label{c25}
\\
& = &  \frac{1}{2} \sum_{j \in B} (R(\vep_j) - \vep_j)
\nonumber
\earray
where we used (\ref{c22}) and $\delta M = - P$. We have to add to
eq.(\ref{c25}) the contribution of the single occupied levels,
namely $\frac{1}{2} \sum_{j \in B} \vep_j$ which then coincides with
the Bogoliubov formula (\ref{c21}). This proves the one to one
correspondence between  the Bogoliubov quasiparticles occupying
different energy levels and the blocked levels in the canonical
ensemble.

\end{document}